\documentclass[times,twocolumn,final]{elsarticle}
\usepackage{medima}
\usepackage{framed,multirow}

\usepackage{amssymb}
\usepackage{latexsym}

\usepackage{url}
\usepackage{xcolor}

\usepackage{hyperref}
\usepackage{float}

\definecolor{newcolor}{rgb}{.8,.349,.1}

\journal{Computerized Medical Imaging and Graphics}

\usepackage[utf8]{inputenc}

\begin{document}

\verso{Taro Langner \textit{et~al.}}

\begin{frontmatter}
	
	\title{Uncertainty-Aware Body Composition Analysis with Deep Regression Ensembles \newline on UK Biobank MRI}%
	%\tnotetext[tnote1]{Funding: This work was supported by a research grant from the Swedish Heart- Lung Foundation and the Swedish Research Council (2016-01040, 2019-04756, 2020-0500, 2021-70492).}
	
	\author[1]{Taro \snm{Langner}\corref{cor1}
	\cortext[cor1]{
		Corresponding author. \\ E-mail address: taro.langner@surgsci.uu.se (T. Langner)%\ead{taro.langner@surgsci.uu.se}
		}	
	}	
	\author[2]{Fredrik K. \snm{Gustafsson}}
	\author[3]{Benny \snm{Avelin}}
	\author[2]{Robin \snm{Strand}}
	\author[1,4]{H\r{a}kan \snm{Ahlstr\"{o}m}}	
	\author[1,4]{Joel \snm{Kullberg}}

	\address[1]{Uppsala University, Department of Surgical Sciences, Uppsala, Sweden}
	\address[2]{Uppsala University, Department of Information Technology, Uppsala, Sweden}
	\address[3]{Uppsala University, Department of Mathematics, Uppsala, Sweden}
	\address[4]{Antaros Medical AB, BioVenture Hub, M\"{o}lndal, Sweden}
	
	%\received{1 May 2013}
	%\finalform{10 May 2013}
	%\accepted{13 May 2013}
	%\availableonline{15 May 2013}
	%\communicated{S. Sarkar}

\begin{abstract}
Along with rich health-related metadata, medical images have been acquired for over 40,000 male and female UK Biobank participants, aged 44-82, since 2014. Phenotypes derived from these images, such as measurements of body composition from MRI, can reveal new links between genetics, cardiovascular disease, and metabolic conditions. 
In this work, six measurements of body composition and adipose tissues were automatically estimated by image-based, deep regression with ResNet50 neural networks from neck-to-knee body MRI. Despite the potential for high speed and accuracy, these networks produce no output segmentations that could indicate the reliability of individual measurements. The presented experiments therefore examine uncertainty quantification with mean-variance regression and ensembling to estimate individual measurement errors and thereby identify potential outliers, anomalies, and other failure cases automatically. 
In 10-fold cross-validation on data of about 8,500 subjects, mean-variance regression and ensembling showed complementary benefits, reducing the mean absolute error across all predictions by 12\%. Both improved the calibration of uncertainties and their ability to identify high prediction errors. With intra-class correlation coefficients (ICC) above 0.97, all targets except the liver fat content yielded relative measurement errors below 5\%. Testing on another 1,000 subjects showed consistent performance, and the method was finally deployed for inference to 30,000 subjects with missing reference values. The results indicate that deep regression ensembles could ultimately provide automated, uncertainty-aware measurements of body composition for more than 120,000 UK Biobank neck-to-knee body MRI that are to be acquired within the coming years.

\end{abstract}


\end{frontmatter}

\section{Introduction}

UK Biobank studies more than half a million volunteers by collecting data on blood biochemistry, genetics, questionnaires on lifestyle, and medical records \citep{sudlow_uk_2015}. 

For 100,000 participants, the ongoing examinations also include medical imaging, such as dedicated MRI of the brain, heart, liver, pancreas, and the entire body from neck to knee \citep{littlejohns2020uk}. Ongoing repeat imaging for 70,000 subjects will furthermore enable longitudinal studies over two or more years. Image-derived phenotypes, such as measurements of body composition and organ volumes, hold the potential for non-invasive studies of aging, cardiovascular, and metabolic conditions at large scale within this cohort. 

The relationship between obesity, type-2 diabetes, and nonalcoholic fatty liver disease is of particular interest due to their high prevalence and associated adverse health effects \citep{wilman2017characterisation, linge2018body}. Depending on genetic and environmental factors, body fat can accumulate in organs, abdominal depots, and muscle infiltrations, all of which have specific effects on health outcomes. Ongoing work is therefore concerned with acquiring measurements of liver fat content \citep{wilman2017characterisation}, muscle volumes, and adipose tissue depots \citep{west_feasibility_2016, linge2018body} with manual and semi-automated techniques \citep{borga2018mri}. Recent works also proposed fully-automated techniques with neural networks for segmentation, which have been applied to the heart \citep{bai2018automated}, kidney \citep{langner2020kidney}, pancreas \citep{basty2020automated,bagur2020pancreas}, and liver \citep{irving2017deep}, but also the iliopsoas muscles \citep{fitzpatrick2020large}, spleen, adipose tissues, and more \citep{liu2020systematic}. 
Similar to the latter, neural networks have also been proposed for segmentation of adipose tissues in other studies involving computed tomography (CT) \citep{wang2017two, weston2019automated} and MRI \citep{langner2019fully, estrada2020fatsegnet, kustner2020fully}.

Apart from semantic segmentation, neural networks can also be trained for image-based regression, predicting numerical measurement values without any need for explicit delineations. In medical imaging, deep regression has gained attention for analyses of human age in MRI of the brain \citep{cole2018brain}, volume measurements of the heart \citep{xue2017direct}, and blood pressure, sex, and age in retinal fundus photographs \citep{poplin2018prediction}. On UK Biobank neck-to-knee body MRI, deep regression can quantify human age and liver fat, but also various measurements of body composition. For the latter, its accuracy can exceed the agreement between established gold standard techniques \citep{Langner2020}.

This type of deep regression requires no ground truth segmentations and can measure abstract properties by training on numerical reference values from arbitrary sources.
However, the lack of output segmentations poses a limitation, as the predicted numerical values give no indication of confidence or reliability. Previous work examined the underlying relevant image features with saliency analysis, but only provided interpretations on cohort level without attempting to estimate individual measurement errors. 

Recent advances in the field of uncertainty quantification have the potential to address some of these concerns by providing an error estimate for each individual measurement \citep{ghahramani2015probabilistic}. High uncertainty could accordingly alert researchers or clinical operators to anomalies, outliers, or other failure cases of these systems \citep{kendall2017uncertainties}. Among various proposed methods, such as Bayesian inference with Markov chain Monte-Carlo techniques \citep{neal2012bayesian} and more computationally viable approximations that apply dropout at test time \citep{gal2016dropout}, recent work reported superior behavior for deep ensembling strategies \citep{gustafsson2020evaluating, ovadia2019can, ashukha2020pitfalls}. These approaches provide \textit{predictive uncertainty} by training multiple neural networks to each predict not only a point estimate but a probability distribution, with multiple network instances forming an ensemble \citep{lakshminarayanan2017simple}. In related work, a similar approach was recently applied for age estimation from fetal brain MRI, reporting high accuracy and promising indications for abnormality detection \citep{shi2020fetal}.

The aim of this work is to develop an automated strategy for body composition analysis on UK Biobank neck-to-knee body MRI which provides not only measurements \citep{Langner2020} but also introduces individual uncertainty estimates that can represent confidence intervals. As a key advantage, the deep regression approach can be trained without access to reference segmentation masks and instead learns to emulate the existing, numerical metadata. Six body composition measurements relating to adipose tissues with high relevance for cardiometabolic disease were predicted from two-dimensional representations of the MRI data. ResNet50 neural network instances \citep{he_deep_2016} for image-based regression were trained to each predict the mean and variance of a Gaussian probability distribution over a given measurement value. Combined into ensembles they provided estimates of predictive uncertainty \citep{lakshminarayanan2017simple}. 
The main contribution consists in extensive analysis of the independent effects of \textit{mean-variance} regression and ensembling on overall accuracy and speed, but also on the calibration \citep{guo2017calibration} of uncertainties and their ability to identify the worst predictions in sparsification \citep{ilg2018uncertainty}, both in cross-validation on about 8,500 subjects and testing on another 1,000 subjects. The proposed method was deployed for inference to obtain previously unavailable measurements from more than 30,000 images, including 1,000 repeat scans.

\section{Materials and methods}

The neck-to-knee body MRI of each subject was formatted into a two-dimensional image from which the proposed method estimates a numerical measurement value in image-based regression. This work examines \textit{least squares} regression, which produces only the measurement value itself, \citep{Langner2020,langner2020large}, but also \textit{mean-variance} regression \citep{nix1994estimating}, in which both the mean value and the variance of a Gaussian probability distribution over one measurement of one subject is modeled. In \textit{ensembling}, the predictions of several networks are furthermore aggregated \citep{lakshminarayanan2017simple}. The thus obtained uncertainty estimates can help to identify outliers and potential failure cases automatically \citep{gustafsson2020energy}.

\begin{figure*}[h!]	
	\centering	
	\includegraphics[width=\textwidth]{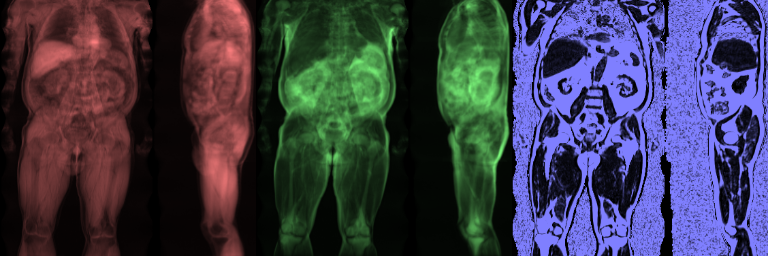}	
	\caption{As input to the neural network, each MRI volume was represented as color image of $(256 \times 256 \times 3)$ pixels by forming channels from the projected water (red) and fat (green) signal and fat fraction slices (blue) from two axes each.}
	\label{fig_input}		
\end{figure*}

\subsection{UK Biobank image data}

UK Biobank has recruited more than half a million men and women by letter from the National Health Service in the United Kingdom, starting in 2006 \citep{sudlow_uk_2015}. Examinations involve several visits to UK Biobank assessment centers, with imaging procedures launching in 2014 for a subgroup of 100,000 participants \citep{littlejohns2020uk}. At the time of writing, medical imaging data from three different centers has been released for 40,264 men and women (52\% female) aged 44-82 (mean 64) years with BMI 14-62 (mean 27) $kg / m^2$ and a majority of 94\% with self-reported White-British ethnicity. 

For 1,209 of these, data from a repeat imaging visit with an offset of about two years has been released. All participants provided informed consent and both the UK Biobank examinations and the experiments in this work were approved by the responsible British and Swedish ethics committees.

\subsubsection{MRI data}

The MRI protocol examined in this work is listed as UK Biobank field 20201 and covers the body from neck to knee in six separate imaging stations acquired in a scan time below ten minutes \citep{west_feasibility_2016,littlejohns2020uk}. Volumetric, co-aligned images of water and fat signal were acquired with a two-point Dixon technique with TR = 6.69, TE = 2.39/4.77 ms and flip angle 10deg on a Siemens Aera Magnetom 1.5 device. The image resolution varies between stations, with a typical grid of \mbox{($224 \times 174 \times 44)$ voxels} of \mbox{$(2.232 \times 2.232 \times 4.5)$ mm} (for more detail, see "Body MRI protocol parameters" in \cite{littlejohns2020uk}).

\subsubsection{Image formatting}

For this work, the six MRI stations of each subject were first fused into a common voxel grid by trilinear interpolation to form a single volume of \mbox{$(224 \times 174 \times 370)$ voxels} for each signal type. These volumes were then converted to two-dimensional representations by summing all values along two axes of view, yielding a coronal and sagittal mean intensity projection, which were concatenated side by side. This was done separately for both the water and fat signal, with the resulting images individually normalized and downsampled to form two color channels of a single image of \mbox{$(256 \times 256 \times 2)$} pixels \citep{Langner2020}. As a third image channel, both a single coronal and sagittal fat fraction slice were extracted based on a body mask \citep{langner2020large}. These fractions resulted from voxel-wise division of the fat signal by the sum of water and fat signal. Fig. \ref{fig_input} shows the result, a dual mean intensity projection with fat fraction slices, encoded in 8bit for faster processing.

\subsection{Ground truth}

UK Biobank provides several body composition measurements from the same neck-to-knee body MRI data as used in this work, based on volumetric multi-atlas segmentations \citep{west_feasibility_2016,borga2015validation}: Visceral Adipose Tissue (VAT), abdominal Subcutaneous Adipose Tissue (SAT), Total Adipose Tissue (TAT), Total Lean Tissue (TLT), and Total Thigh Muscle (TTM). Together with Liver Fat Fraction (LFF) values based on dedicated multi-echo liver MRI \citep{linge2018body}, these reference measurements form the ground truth data, or regression targets, for this work.

\subsection{Data partitions}

Among the 40,264 released images of the initial imaging visit, visual inspection identified 1,376 subjects with artifacts such as water-fat signal swaps, non-standard positioning and metal objects \citep{Langner2020}. 
Three datasets were formed from the initial imaging visit from those subjects for whom any of the six reference measurements were available.

Dataset \textit{$D_{cv}$} consists of 8,539 subjects without artifacts and was subdivided into a 10-fold cross-validation split which was retained for all experiments.

Dataset \textit{$D_{test}$} contains another 1,107 subjects without artifacts and served as a test set, but notably lacks any values for two of the six regression targets for which no reference values have been released yet.

Dataset \textit{$D_{art}$} was formed from those subjects with identified artifacts, yielding 330 subjects, to examine behavior on abnormal data. 

Two additional datasets were formed from those subjects with no available reference measurements.
Dataset \textit{$D_{infer}$} comprises all remaining 29,234 subjects without artifacts from the initial imaging visit, for whom the prediction model was applied to for inference.
Finally, dataset \textit{$D_{revisit}$} was formed for inference on the repeat imaging visit from 1,179 subjects with no image artifacts.

\subsection{Model}

A ResNet50 architecture \citep{he_deep_2016} was configured to receive the two-dimensional image format as seen in Fig. \ref{fig_input} as input for a given subject and predict all six regression targets at once. No explicit segmentation was performed at any stage of this work. Each network was pre-trained on ImageNet and optimized with Adam \citep{kingma2014adam} at batch size 32 with online augmentation by random translations. After 5,000 iterations, the base learning rate of 0.0001 was reduced by factor 10 and training continued for another 1,000 iterations \citep{Langner2020}. All experiments were conducted in PyTorch, using an Nvidia RTX 2080 Ti graphics card with 11GB RAM.
%\vspace{1cm}

Four distinct configurations were compared. As the first one, a \textit{least squares} regression network predicted only these six output values, each corresponding to one measurement for a given subject, trained by optimizing the mean squared error criterion of equation \ref{eq_mse}. In this formula, $\mu_\theta(\mathbf{x}_n)$ represents the network prediction for the $n$-th input sample $\mathbf{x}_n$, with $y_n$ as the corresponding ground truth value. 

\begin{equation}
MSE = \frac{1}{N}\sum_{n=1}^{N}(y_n - \mu_\theta(\mathbf{x}_n))^2
\label{eq_mse}
\end{equation}

As a second configuration, \textit{least squares ensembles} were formed by combining ten such networks. Their predictions were averaged and the spread, or empirical variance, of their predictions used as uncertainty estimate \citep{ilg2018uncertainty}. 

As the third configuration, \textit{mean-variance} regression was performed by predicting two values, corresponding to the mean and variance of a Gaussian probability distribution over one measurement value for a given subject, optimized with a negative log-likelihood criterion \citep{nix1994estimating} as shown in equation \ref{eq_nll}. Here, $p_\theta(y_n | \mathbf{x}_n)$ is the probabilistic predictive distribution over one measurement value, modeled by the network outputs $\mu_\theta(\mathbf{x}_n)$ and $\sigma^2_\theta(\mathbf{x}_n)$, which represent the predicted mean and corresponding predicted variance for input sample $\mathbf{x}_n$, respectively. The last term, $c$, is a constant that does not depend on $\theta$. This criterion expands the mean squared error of eq. \ref{eq_mse} by a sample-specific, heteroscedastic variance and can likewise be averaged across multiple samples. This predicted variance directly serves as an estimate of uncertainty, with high values describing a wide normal distribution within which plausible values for the estimated measurement are assumed.

\begin{equation}
-\log p_\theta(y_n | \mathbf{x}_n) = \frac{\log \sigma^2_\theta(\mathbf{x}_n)}{2} + \frac{(y_n - \mu_\theta(\mathbf{x}_n))^2}{2\sigma^2_\theta(\mathbf{x}_n)} + c
\label{eq_nll}
\end{equation}

As the fourth and final configuration, \textit{mean-variance ensembles} employ ten such network instances. Their predictions can likewise be aggregated to obtain estimates of predictive uncertainty \citep{lakshminarayanan2017simple}. 

In all ensembles, model diversity was increased by withholding one of ten evenly sized subsets of the training data from each instance, as if they had been obtained from a preceding cross-validation experiment.
The target values were standardized \citep{Langner2020}. When one or more of the six ground truth values for a given training sample were missing, their contribution to the loss term was dynamically set to zero, so that they would not affect the training process. In this way, it was possible to utilize samples with missing values and provide as much training data as possible. A PyTorch implementation for training and inference will be made publicly available\footnote{github.com/tarolangner/ukb\_mimir}.

\subsection{Evaluation}

All configurations were evaluated in 10-fold cross-validation on dataset \textit{$D_{cv}$} and also validated against artifact dataset \textit{$D_{art}$}. The best configuration was eventually applied to test dataset \textit{$D_{test}$} and deployed for inference on datasets \textit{$D_{infer}$} and \textit{$D_{revisit}$}.

The predicted measurements were compared to the reference values with the intraclass correlation coefficient (ICC) with a two-way random, single measures, absolute agreement definition \citep{koo2016guideline} and the coefficient of determination R$^2$. The mean absolute error (MAE) is also reported, together with the mean absolute percentage error (MAPE) as a relative error measurement. The latter is the absolute difference between prediction and reference divided by the reference. Additionally, aggregated saliency maps were generated to highlight relevant image areas \citep{selvaraju_grad-cam:_2017}.

The estimated uncertainties were evaluated regarding sparsification \citep{ilg2018uncertainty} and calibration \citep{guo2017calibration}. 
Sparsification examines whether the highest uncertainties coincide with the highest prediction errors. Ranking all measurements by their uncertainty and excluding one after another should accordingly yield consistent improvements in performance metrics such as the MAE. 
Calibration examines the magnitude of uncertainties and resulting under- or overconfidence of predictions. The uncertainty obtained for any given sample corresponds to the variance of a Gaussian probability distribution, modeling characteristic confidence intervals around the predicted mean. Higher uncertainty scales these intervals to be wider, enabling them to cover larger errors. Ideally calibrated uncertainties define confidence intervals that cover, on a set of samples, a percentage of errors that corresponds exactly to their specific confidence level.

\section{Results}

\begin{figure*}[h!]	
	\centering	
	\includegraphics[width=\textwidth]{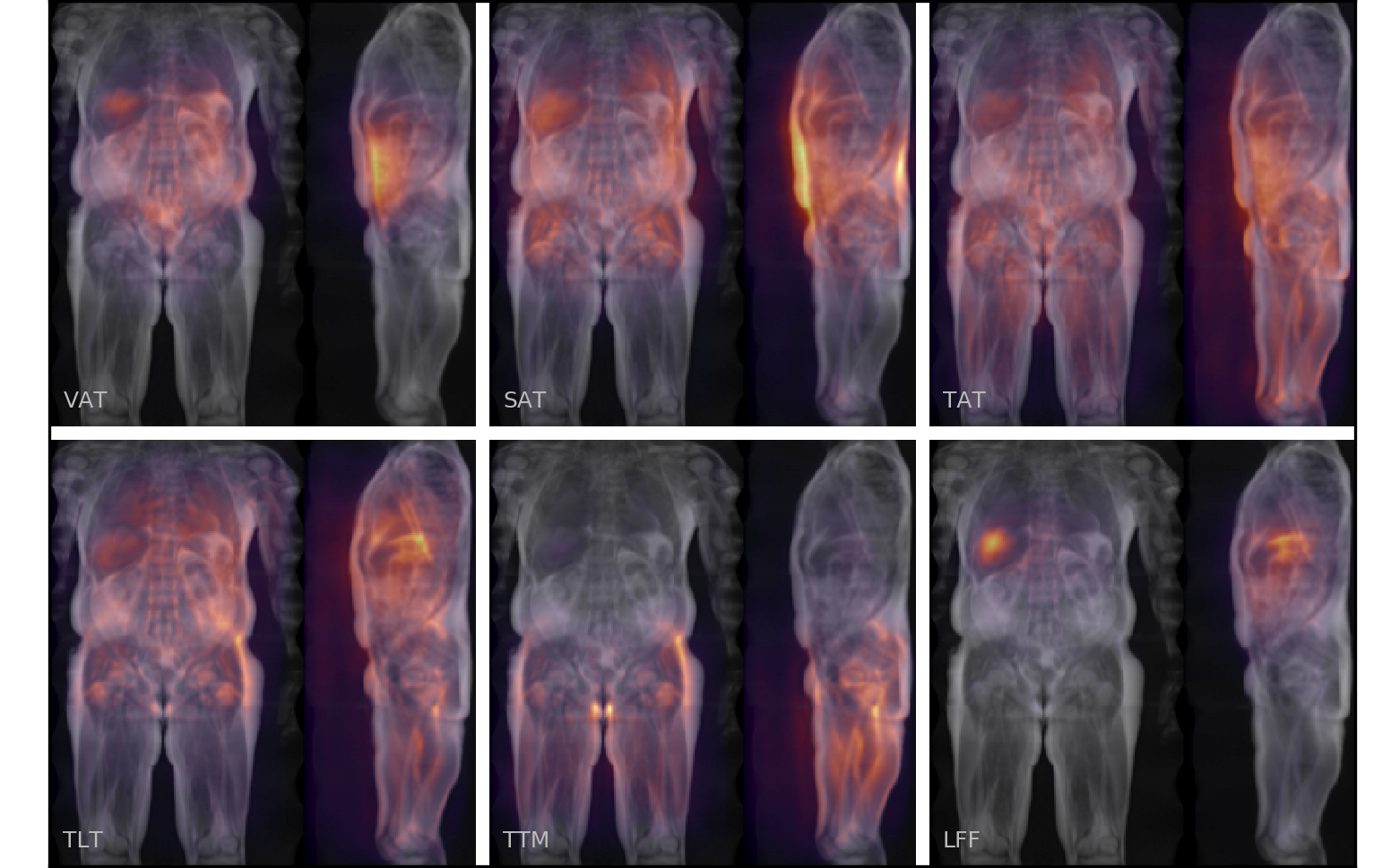}	
	\caption{Co-registered, aggregated saliency information for about 3,000 subjects, showing the fat signal channel only (see supplementary material for more). }
	\label{fig_saliency}		
\end{figure*}

Both \textit{mean-variance} regression and ensembling provided complementary benefits. Combining both yielded the best predictive performance, shown in Table \ref{tab_evaluation} and Fig.~\ref{fig_scatter}, with additional detail provided in the supplementary material. On average, the predictions can account for 98\% (R$^2$) of the variability in reference values, with absolute agreement (ICC) above 0.97 on all targets. The metrics carry over to the test data largely unchanged. All targets are predicted with a relative error below 5\%, except the liver fat fraction. This target also incurred the highest relative uncertainties and is examined further in the supplementary material, together with additional evaluation metrics, and a comparison to alternative reference methods. It also provides additional detail on the saliency analysis, which is compiled into Fig.~\ref{fig_saliency}.

\begin{figure*}[h]%[H]
	
	\centering	
	
	\includegraphics[width=\textwidth]{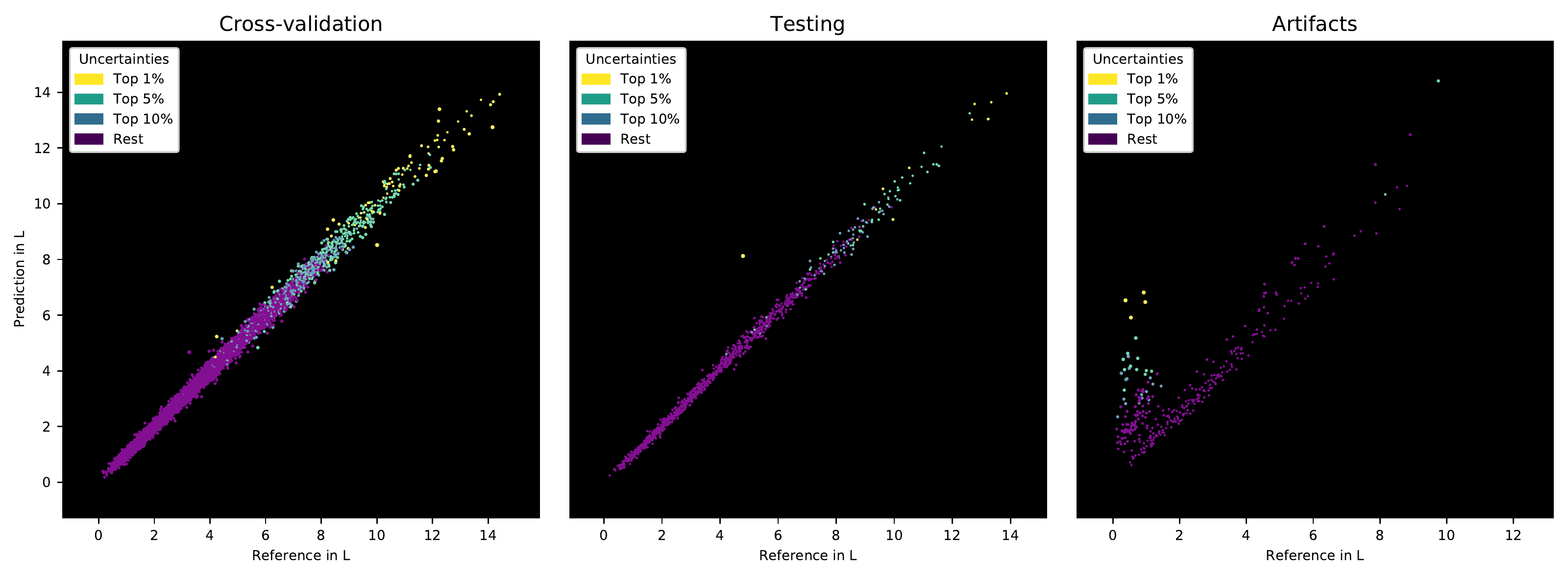}		
	
	\caption{Mean-variance ensemble predictions and reference values for Visceral Adipose Tissue (VAT) in cross-validation on \textit{$D_{cv}$}, testing on \textit{$D_{test}$}, and on subjects with artifacts of \textit{$D_{art}$}, depicted with color-coded uncertainty. The listed percentiles refer to those samples with the highest uncertainty.}
	
	\label{fig_scatter}	
	
\end{figure*}

\begin{table*}
	
	\begin{center}	
		
		\renewcommand{\arraystretch}{1.25}	
		
		\caption{Evaluation results}	
		
		\label{tab_evaluation}		
		
		\begin{tabular}{lc|r@{\hskip 0.8cm}r@{\hskip 0.5cm}|@{\hskip 0.5cm}rr}
			
			\hline	
			
			\hline	
			
			&&\multicolumn{2}{c}{Cross-Validation } & \multicolumn{2}{c}{Testing} \\
			
			Target name &   & ICC & \% error   &   ICC & \% error\\
			
			\hline			
			
			Visceral Adipose Tissue & (VAT) &   0.997  & 4.2  & 0.997 & 3.6 \\
			
			Abdominal Subcutaneous Adipose Tissue & (SAT) &  0.996  & 2.8  & 0.996 & 2.7 \\			
			
			Total Adipose Tissue & (TAT)&   0.997  & 1.8  & /     & / \\
			
			Total Lean Tissue & (TLT) &   0.983  & 2.5  & /     & / \\
			
			Total Thigh Muscle & (TTM) &   0.996  & 1.6  & 0.995 & 1.6 \\			
			
			Liver Fat Fraction & (LFF) & 0.979 & 25.7 & 0.982 & 21.6 \\
			
			\hline	
			
			\hline	
			
			\multicolumn{6}{l}{* Results for the \textit{mean-variance} ensemble on cross-validation dataset \textit{$D_{cv}$} and testing on dataset \textit{$D_{test}$}, } \\
			
			\multicolumn{6}{l}{  with intraclass correlation coefficient (ICC) and MAPE (\% error).} \\
			
		\end{tabular}
	\end{center}
\end{table*}

\begin{figure}%[H]
	
	\centering	
	
	\includegraphics[width=0.5\textwidth]{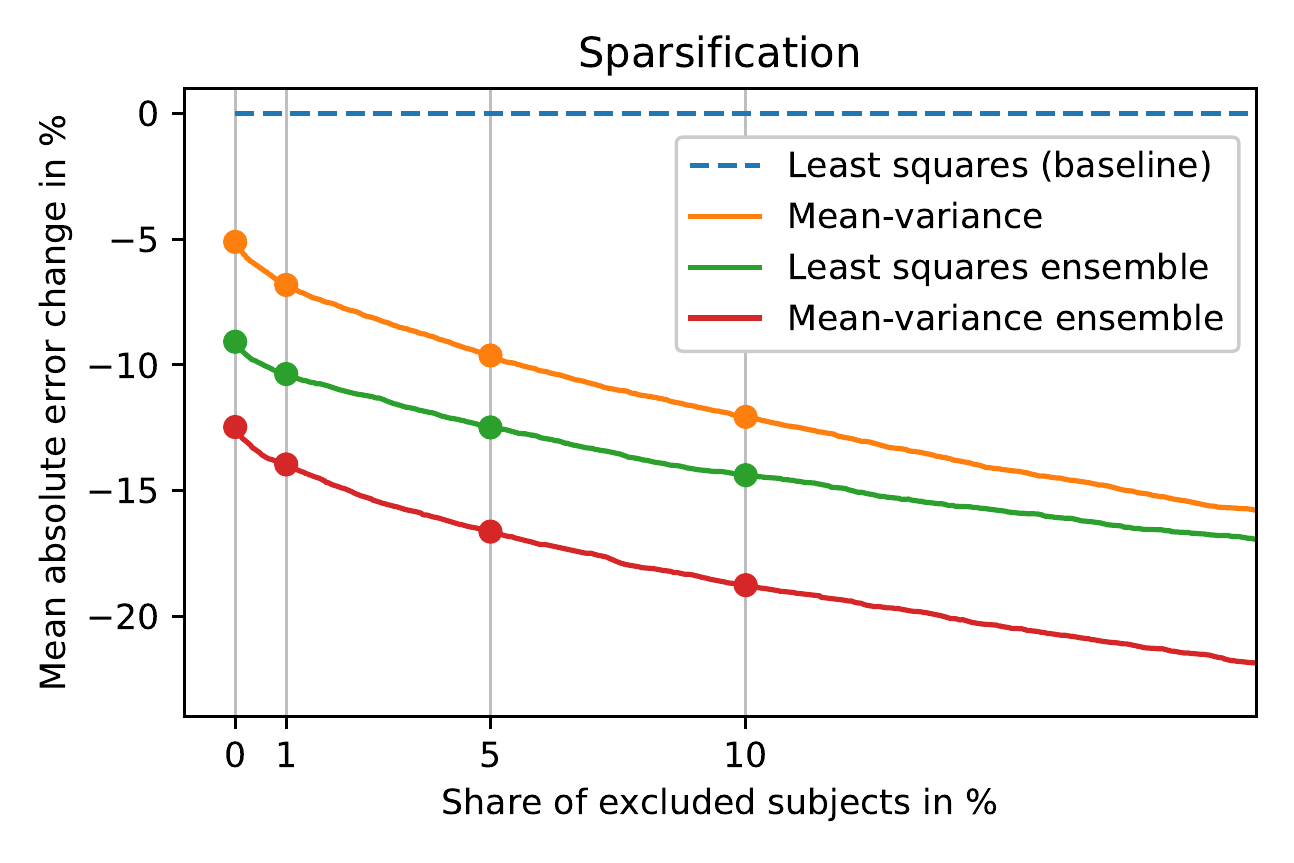}		
	
	\caption{This sparsification plot \citep{ilg2018uncertainty} shows how the overall performance can be improved by gradually excluding those subjects with the highest predicted measurement uncertainty. Each position along the x-axis represents a certain share of excluded, most uncertain measurements, whereas the y-axis shows the change in mean absolute error relative to baseline, averaged across all targets on dataset \textit{$D_{cv}$}. Even without utilizing the uncertainty to exclude any subjects, the \textit{mean-variance} ensemble achieves a reduction of the MAE by 12\%. Further improvements in the MAE can be achieved excluding increasingly large shares of those measurements with highest uncertainty.}
	
	\label{fig_spars_curves}

\end{figure}

Fig. \ref{fig_spars_curves} shows that even without utilizing the uncertainties, the \textit{mean-variance} regression ensemble reduces the MAE by 12\% when compared to the \textit{least-squares} regression baseline. The uncertainties enable sparsification, identifying some of the worst predictions which can be excluded to reduce the prediction error even further. The scatter plots of Fig. \ref{fig_scatter} show predictions for one target in detail, together with color-coded uncertainty.
Despite containing image artifacts, not all subjects of dataset \textit{$D_{art}$} yield higher uncertainties than the normal material. Indeed, many of these subjects result in highly accurate predictions despite the artifacts, and high uncertainties tend to occur only in those cases with high prediction errors.
On test dataset \textit{$D_{test}$}, the uncertainty highlights an outlier case for VAT (see Fig. \ref{fig_scatter}), SAT, and TTM. This one subject causes consistently flawed predictions and was found to suffer from an abnormal, atrophied right leg.

On datasets \textit{$D_{cv}$} and \textit{$D_{test}$} the predicted means exhibit a consistent, linear correlation with the predicted log uncertainties. Accordingly, large subjects with high volumes induce systematically higher uncertainty. Although these cases also generally incur higher prediction errors, this bias can be shown to not achieve optimal sparsification. On the normal material with hardly any outliers, this tendency is so strong that sparsifying simply by predicted mean is almost as effective as using the uncertainties. On dataset \textit{$D_{art}$}, this bias is less pronounced, as those cases with artifacts that cause genuine prediction failures are correctly assigned much higher uncertainty.

The best calibration was also achieved by the \textit{mean-variance} ensemble, which nonetheless often produced overconfident uncertainties. Post-processing with target-wise scaling factors can achieve a near perfect fit to the validation data, however, and also improves the overall calibration on the test set. The supplementary material explores both sparsification and calibration in more detail and also lists results for datasets \textit{$D_{infer}$} and \textit{$D_{revisit}$}, on which the proposed method inferred new measurements for over 30,000 images.

No difference in processing speed was observed between \textit{least squares} and \textit{mean-variance} regression. Image formatting required the bulk of processing time, but once cached, training one network only requires about 15 minutes, or 2.5 hours for an ensemble of ten instances. Ensemble predictions for about 60 subjects can be generated within one second, so that inference for all 30,000 required less than ten minutes.

\section{Discussion}

% PREDICTIVE PERFORMANCE

With relative measurement errors below 5\%, all targets except the liver fat fraction can be predicted with higher accuracy than observed for the mutual agreement between the reference and alternative established methods, both in cross-validation and on the test data. For liver fat itself, the relative error of 22-26\% is worse than the 15\% seen between the reference used here and an alternative set of UK Biobank liver fat measurements. The two-point Dixon images inherently limit the prediction accuracy for this target, as the reference values were obtained from another imaging protocol that reconstructs fat fractions more faithfully \citep{wilman2017characterisation, linge2018body}. The saliency analysis of Fig.~\ref{fig_saliency} indicates that the networks nonetheless learned to correctly identify liver tissue and other target-specific regions. The inference on 30,000 subjects provides material for further medical study which is, however, beyond the scope of this work.

% UNCERTAINTY

The estimated uncertainties identified many of the worst prediction errors. They correctly highlighted an outlier with abnormal physiology on the test data and enabled consistent reductions in the mean prediction error by excluding the least certain measurements. On the inference datasets, the highest uncertainties were furthermore found in several cases to coincide with previously undetected anomalies in positioning, but also with minor artifacts and pathologies that may have negatively affected prediction accuracy and should arguably have been excluded during the original quality controls. In practice, the acquired measurements can accordingly be supplied together with their uncertainty, which could serve both as an error estimate and as a means to identify potential anomalies and failure cases. The affected cases could then be manually examined and, if necessary, excluded from further analyses.

However, the results also show two noteworthy limitations of the proposed approach which arise from imperfect calibration and the observed bias for high measurement values to incur high uncertainties. The imperfect calibration is linked to uncertainties that often underestimate the true measurement error. This is a known effect related to overfitting on the training data \citep{guo2017calibration, laves2020well}. As shown in the supplementary material, it is possible to correct the calibration by calculating target-wise scaling factors on the validation results. Once obtained, these simple scaling factors also yield improved overall calibration on the test data.

% UNCERTAINTY BIAS

The bias towards systematically higher uncertainty in higher measurement values is a more concerning pattern. This effect can make it hard to distinguish whether a measurement with high uncertainty should be excluded due to being flawed or whether it merely resulted from a large subject, many of whom may provide valuable insight in correlation studies. It is most pronounced in the normal material where no genuine failure cases are encountered. In contrast, the uncertainty for one abnormal subject in the test set or the flawed predictions on images with artifacts of dataset \textit{$D_{art}$} are typically higher.

Conceptually, body weights above 150kg and BMIs of up to 53 kg/m$^2$ as present in the training data represent physiological extremes that could be considered outliers in their own right.
Arguably, the two-dimensional projections are also inherently less suitable to represent more voluminous bodies and many of the largest subjects furthermore show considerable variability in shape and extend beyond the field of view. Even then, the effect is gradual and large subjects incur higher uncertainty than warranted in terms of the prediction errors alone. Previous work on age estimation from fetal brain MRI reported similar effects \citep{shi2020fetal}, noting specifically that higher aleatoric uncertainty, corresponding to the variances returned by the network instances, correlated with higher gestational age of the fetal brain. In this work, the effect is present in both the aleatoric and epistemic uncertainty component as modeled by the empirical variance, even in \textit{least-squares} regression ensembles. %The origin and possible solutions for this bias remain to be explored in future work, but it appears that this bias can not be trivially resolved by inverting the magnitudes of target values by simple changing their sign before and after the calculations.

% TECHNICAL
On a technical level, the \textit{mean-variance} configuration provided immediate benefits over \textit{least squares} regression despite merely changing the loss function and requiring that both a mean and a variance be predicted. This could be explained by \textit{loss attenuation} \citep{kendall2017uncertainties,ilg2018uncertainty} weakening the impact of outliers among the ground truth values. Several mismatches between the image data and reference were identified where the predictions also incur high errors in spite of low uncertainty. Images with artifacts, in contrast, did not necessarily yield high uncertainties, as the method was in fact able to provide accurate predictions for many of them. In turn, this also means that subjects with artifacts will not generally be identified as out-of-distribution samples. Ensembling yielded an inherent benefit in prediction accuracy and also improved the calibration. The ten network instances were conveniently obtained from a cross-validation split, but sufficient ensemble diversity could potentially be induced by random weight initialization alone and similar benefits can be achieved with fewer instances as seen in ablation experiments of the supplementary material and related literature \citep{fort2019deep, ovadia2019can}. Based on the results, even a single \textit{mean-variance} instance would be viable in practical settings if model size and runtime are of chief concern. The calibration could be adjusted with scaling factors, although it would not benefit from the 12\% reduction in MAE achieved by ensembling.

% LIMITATIONS
Several additional limitations apply on a methodological level. No independent, external test set was examined, so that no claim can be made about generalization of the trained networks to other studies. The validation and test cases used in this work are furthermore preselected for the intended measurements by virtue of having passed the quality controls of the reference methods. Similarly, certain phenotypes were systematically excluded from the experiments in this paper, such as subjects with knee implants or other severe pathologies. When applied to different imaging devices, protocols, or subject demographics, new training data in the range of several hundred samples would likely be required. In contrast, multi-atlas segmentations with manual corrections have been based on just above 30 annotated subjects \citep{west_feasibility_2016}, whereas neural networks for semantic segmentation typically report training data ranging from 90 to 220 subjects \citep{fitzpatrick2020large, bagur2020pancreas} on UK Biobank MRI. 

When compared to neural networks for segmentation, the proposed approach accordingly requires more training samples and produces no output segmentation masks. 
In turn, it can be trained without access to reference segmentations in an end-to-end fashion that does not require for the property of interest to be manually encoded in the input data during training. Previous work showed that it outperformed segmentation in estimating liver fat from the two-point Dixon images, possibly by using additional image information that is not easily accessible to human intuition \cite{langner2020large}, and also accurately estimated other, more abstract properties \cite{Langner2020}. Likewise, the uncertainty quantification as proposed here can provide error bounds for the measurement that is ultimately of interest for medical research, although approaches for voxel-wise uncertainty from segmentation networks have also been proposed in the literature \cite{roy2019bayesian}.

The concept of designing two-dimensional input formats resembles hand-crafted feature selection and it would be preferable to apply a regression technique directly to the volumetric MRI data. No claim is intended for the chosen representation to be optimal as input to the neural network. The MRI volumes could be sliced, projected, or aggregated in various ways and in any signal or phase component may contain valuable information.  Despite the empirical success of the presented approach, further improvements may be possible, as the chosen format compresses the MRI data to just $0.5\%$ of its original size and almost certainly results in a loss of information. However, a fully volumetric approach would likely require substantially increased processing time and GPU memory. The proposed approach, in contrast, can run on consumer-grade hardware and achieves relative errors as low as 1.6\%, which may be hard to improve much further. Future work may adapt the presented approach to the dedicated liver MRI of UK Biobank, with potential for far more accurate liver fat predictions.

Future work may also explore how the bias between high measurements and high uncertainty can be corrected for and could explore alternative strategies which are known to produce substantially distinct estimates of uncertainty \citep{staahl2020evaluation}. However, it is unclear whether Monte-Carlo techniques that employ dropout at test time \citep{gal2016dropout} could reach sufficient predictive performance, whereas more faithful approximations of Bayesian inference with Markov chain Monte-Carlo \citep{neal2012bayesian} may not be computationally viable. Deep ensembles are often reported as one of the most successful strategies \citep{gustafsson2020evaluating, ovadia2019can, ashukha2020pitfalls} and a suitable alternative will have to achieve better calibration and sparsification without sacrificing predictive accuracy or exceeding the computational limitations in order to be competitive.

In a large-scale study such as the UK Biobank the main strengths of the proposed approach can be exploited. Without any need for further guidance, corrections, or intervention, these values can be inferred for the entire imaged study population, both for existing and future imaging data. The resulting measurements can be obtained for further study and quality control months or years before full coverage has been achieved with the reference techniques. In practice, researchers may apply this system to obtain automated measurements for all upcoming 120,000 UK Biobank neck-to-knee body MRI scans yet to be released, and will be alerted to potential prediction failures by the predictive uncertainty. Future developments may also yield comparable systems that could ultimately be integrated into scanner software to provide fully automated analyses for specific imaging protocols.

\section{Conclusion}

In conclusion, both \textit{mean-variance} regression and ensembling provided complementary benefits for the presented task. Without extensive architectural changes or prohibitive increases in computational cost they enabled fast and accurate measurements of body composition for the entire imaged UK Biobank cohort. The predicted uncertainty can, despite the specified limitations, give valuable insight into potential failure cases and will be made available together with the inferred measurements for further medical studies.

\section*{Acknowledgment}
This work was supported by a research grant from the Swedish Heart- Lung Foundation and the Swedish Research Council (2016-01040, 2019-04756, 2020-0500, 2021-70492) and used the UK Biobank Resource under application no. 14237.

\bibliographystyle{model2-names.bst}\biboptions{authoryear}
\bibliography{refs}

\clearpage

%%%%%%%%%%

\onecolumn

\renewcommand{\figurename}{Supplementary Figure}

\renewcommand{\tablename}{Supplementary Table}

\setcounter{page}{1}
\setcounter{section}{1}

\setcounter{table}{0}

\setcounter{figure}{0} 

\setlength\parindent{0pt}

%\section{Supplementary Material}
{\begin{center} \Large Supplementary Material \end{center}}

\textbf{T}he following pages provide additional detail on predictive performance, sparsification, calibration, and inference with the proposed approach. Unless otherwise specified, all listed results were acquired with the configuration that combines both \textit{mean-variance} regression and ensembling. The individual targets are furthermore examined in detail and compared to alternative UK Biobank reference values. A PyTorch implementation for preprocessing, training, and inference with mean-variance regression on the given image data is available online.\\

\textbf{GitHub repository:} 

\url{https://github.com/tarolangner/ukb_mimir} \\

%Note that this version of the supplementary material is abbreviated. Find an extended version with further detail on individual targets and the inference on GitHub.

%\url{https://github.com/tarolangner/mri-biometry/supplementary_material/} \\

\subsection{Datasets and Predictive Performance}

The effective number of samples in the three datasets used for evaluation is listed in Supplementary Table \ref{supp_tab_datasets}, taking into account missing reference values. The inference dataset \textit{$D_{inf}$} furthermore contained 29,234 and the repeat imaging dataset \textit{$D_{revisit}$} another 1,179 unique samples.

\begin{table}[H]

\begin{center}

\renewcommand{\arraystretch}{1.25}

\caption{Number of Subjects per Dataset}

\label{supp_tab_datasets}

\begin{tabular}{clcccc}

\hline

\hline						

Field ID & Target && Cross-validation & Testing & Artifacts \\%\textit{$D_{cv}$} & \textit{$D_{test}$} &  \textit{$D_{art}$} \\

\hline			

22407 & Visceral Adipose Tissue & (VAT) &  8,534 & 1,096 & 327\\% & 29,234 & 1,179\\

22408 & Abdominal Subcutaneous Adipose Tissue & (SAT) &  8,534  & 1,097 & 326\\% & 29,234 & 1,179\\		

22415 & Total Adipose Tissue & (TAT) & 8,270 & 0 & 242\\% & 29,234 & 1,179\\

22416 & Total Lean Tissue & (TLT) & 8,270 & 0 & 242\\% & 29,234 & 1,179\\

22409 & Total Thigh Muscle & (TTM) & 8,478 & 1,038 & 284\\% & 29,234 & 1,179\\

22436 & Liver Fat Fraction & (LFF) & 8,474 & 1,061 &  323\\% & 29,234 & 1,179\\

\hline

\hline

\multicolumn{6}{l}{*UK Biobank Field IDs and number of available subjects with known reference values per target }\\

\multicolumn{6}{l}{ in cross-validation  on dataset \textit{$D_{cv}$}, testing on dataset \textit{$D_{test}$}, and artifact dataset \textit{$D_{art}$}.}

\end{tabular}

\end{center}

\end{table}

Additional documentation for each target is publicly available in the UK Biobank showcase, based on the listed Field IDs: \\
\url{https://biobank.ndph.ox.ac.uk/showcase/search.cgi} \\

Supplementary Table \ref{supp_tab_performance} lists additional evaluation metrics on all targets in cross-validation and testing. The results of all four configurations in cross-validation are listed in Supplementary Table \ref{supp_tab_performance_configs}.

\begin{table}[H]

\begin{center}

\renewcommand{\arraystretch}{1.25}
\setlength{\tabcolsep}{3 pt}

\caption{Predicted Performance in Detail}

\label{supp_tab_performance}

\begin{tabular}{lcc|lrrrr|lrrrr}

\hline

\hline

& & &\multicolumn{5}{c}{Cross-validation} & \multicolumn{5}{c}{Testing} \\

Target && unit & N & ICC & R$^2$ & MAE & MAPE  & N & ICC & R$^2$ & MAE & MAPE \\

\hline

Visceral Adipose Tissue & (VAT) & L & 8,534 & 0.997 & 0.994 & 0.122 &	4.2 &

1,096 & 0.997 & 0.995 & 0.119 & 3.6 \\

Abdominal Subcutaneous Adipose Tissue & (SAT) & L & 8,534 & 0.996 & 0.993 &	0.191 &	2.8 &

1,097 & 0.996 & 0.992 & 0.192 & 2.7\\		

Total Adipose Tissue & (TAT) & L & 8,270 & 0.997 & 0.995 &	0.358 &	1.8 &

0& &&& \\

Total Lean Tissue & (TLT) & L & 8,270 & 0.983 & 0.966 &	0.579 &	2.5 &

0& &&&\\

Total Thigh Muscle & (TTM) & L & 8,478 & 0.996 & 0.993 & 0.162 & 1.6 &

1,038 & 0.995 & 0.990 & 0.174 & 1.6 \\

Liver Fat Fraction & (LFF) & \% & 8,474 & 0.979 & 0.959 & 0.666 &	25.7 &

1,061 & 0.982 & 0.965 & 0.647 & 21.6 \\

\hline

\hline

\multicolumn{13}{l}{* Results for an ensemble of ten \textit{mean-variance} networks in 10-fold cross-validation on dataset \textit{$D_{cv}$} and testing on  dataset \textit{$D_{test}$}. } \\

\multicolumn{13}{l}{ N: Number of subjects, ICC: Intraclass correlation coefficient, R$^2$: Coefficient of determination, MAE: Mean absolute error, } \\

\multicolumn{13}{l}{MAPE: Mean absolute percentage error.} \\

\end{tabular}

\end{center}

\end{table}

\begin{table}[H]

\begin{center}

\renewcommand{\arraystretch}{1.25}

\caption{Comparison of All Configurations in Cross-Validation}

\label{supp_tab_performance_configs}

\begin{tabular}{lcccc}

\hline

\hline

Configuration & ICC & R$^2$ & MAE & MAPE  \\

\hline

\textbf{Visceral Adipose Tissue (VAT) in L} &&&& \\		

Least squares instance & 0.996 & 0.992 & 0.150 & 5.2\\	

Mean-variance instance & 0.997 & 0.993 & 0.134 & 4.6\\	

Least squares ensemble & 0.997 & 0.993 & 0.133 & 4.6\\	

Mean-variance ensemble & 0.997 & 0.994 & 0.122 & 4.2 \\

&&&& \\	

\textbf{Abdominal Subcutaneous Adipose Tissue (SAT) in L} &&&& \\		

Least squares instance & 0.995 & 0.991 & 0.222 & 3.3 \\	

Mean-variance instance & 0.996 & 0.992 & 0.209 & 3.1 \\	

Least squares ensemble & 0.996 & 0.992 & 0.202 & 3.0 \\	

Mean-variance ensemble & 0.996 & 0.993 & 0.191 & 2.8 \\

&&&& \\	

\textbf{Total Adipose Tissue (TAT) in L} &&&& \\		

Least squares instance & 0.997 & 0.993 & 0.420 & 2.1\\	

Mean-variance instance & 0.997 & 0.994 & 0.390 & 1.9\\	

Least squares ensemble & 0.997 & 0.994 & 0.377 & 1.9\\	

Mean-variance ensemble & 0.997 & 0.995 & 0.358 & 1.8 \\

&&&& \\	

\textbf{Total Lean Tissue (TLT) in L} &&&& \\		

Least squares instance & 0.981 & 0.963 & 0.650 & 2.8\\	

Mean-variance instance & 0.981 & 0.962 & 0.632 & 2.7\\	

Least squares ensemble & 0.983 & 0.966 & 0.594 & 2.6\\	

Mean-variance ensemble & 0.983 & 0.966 & 0.579 & 2.5 \\

&&&& \\	

\textbf{Total Thigh Muscle (TTM) in L} &&&& \\		

Least squares instance & 0.996 & 0.991 & 0.182 & 1.8 \\	

Mean-variance instance & 0.996 & 0.992 & 0.176 & 1.8 \\	

Least squares ensemble & 0.996 & 0.993 & 0.163 & 1.6 \\	

Mean-variance ensemble & 0.996 & 0.993 & 0.162 & 1.6 \\

&&&& \\	

\textbf{Liver Fat Fraction (LFF) in \%} &&&& \\		

Least squares instance & 0.977 & 0.956 & 0.706 & 28.1\\	

Mean-variance instance & 0.977 & 0.954 & 0.702 & 26.6\\	

Least squares ensemble & 0.979 & 0.960 & 0.671 & 27.0\\	

Mean-variance ensemble & 0.979 & 0.959 & 0.666 & 25.7 \\

%Abdominal Subcutaneous Adipose Tissue & (SAT) & L & 8,534 & 0.996 & 0.993 &	0.191 &	2.8 \\		

%Total Adipose Tissue & (TAT) & L & 8,270 & 0.997 & 0.995 &	0.358 &	1.8  \\

%Total Lean Tissue & (TLT) & L & 8,270 & 0.983 & 0.966 &	0.579 &	2.5 \\

%Total Thigh Muscle & (TTM) & L & 8,478 & 0.996 & 0.993 & 0.162 & 1.6 \\

%Liver Fat Fraction & (LFF) & \% & 8,474 & 0.979 & 0.959 & 0.666 &	25.7 \\

\hline

\hline

\multicolumn{5}{l}{* Results for all configurations in 10-fold cross-validation on dataset \textit{$D_{cv}$}.} \\

\multicolumn{5}{l}{ N: Number of subjects, ICC: Intraclass correlation coefficient, R$^2$: Coefficient of determination, } \\

\multicolumn{5}{l}{ MAE: Mean absolute error, MAPE: Mean absolute percentage error.} \\

%\multicolumn{13}{l}{} \\

\end{tabular}

\end{center}

\end{table}

\newpage

\subsection{Overall Calibration}

All examined configurations are biased towards overconfidence, consistently underestimating the true prediction errors. The predicted uncertainty should accordingly be scaled up. Suitable target-wise scaling factors can be determined to reach a better calibration on the validation data after training \citep{guo2017calibration, laves2020well}. In this work a simple grid search was used, which resulted in the target-wise scaling factors and the \textit{areas under calibration error curve} (AUCE) \citep{gustafsson2020evaluating} shown in Supplementary Table \ref{supp_tab_calibration}, with calibration plots, or reliability diagrams, shown in Supplementary Fig. \ref{supp_fig_calibration_fix}. The same factors also achieve a considerable improvement when applied to the test data, indicating that the calibration of the proposed method could easily be corrected with this strategy for the normal material of the entire cohort.

\vspace{-0.25cm}

\begin{figure}[H]
\centering	
\includegraphics[width=0.666\textwidth]{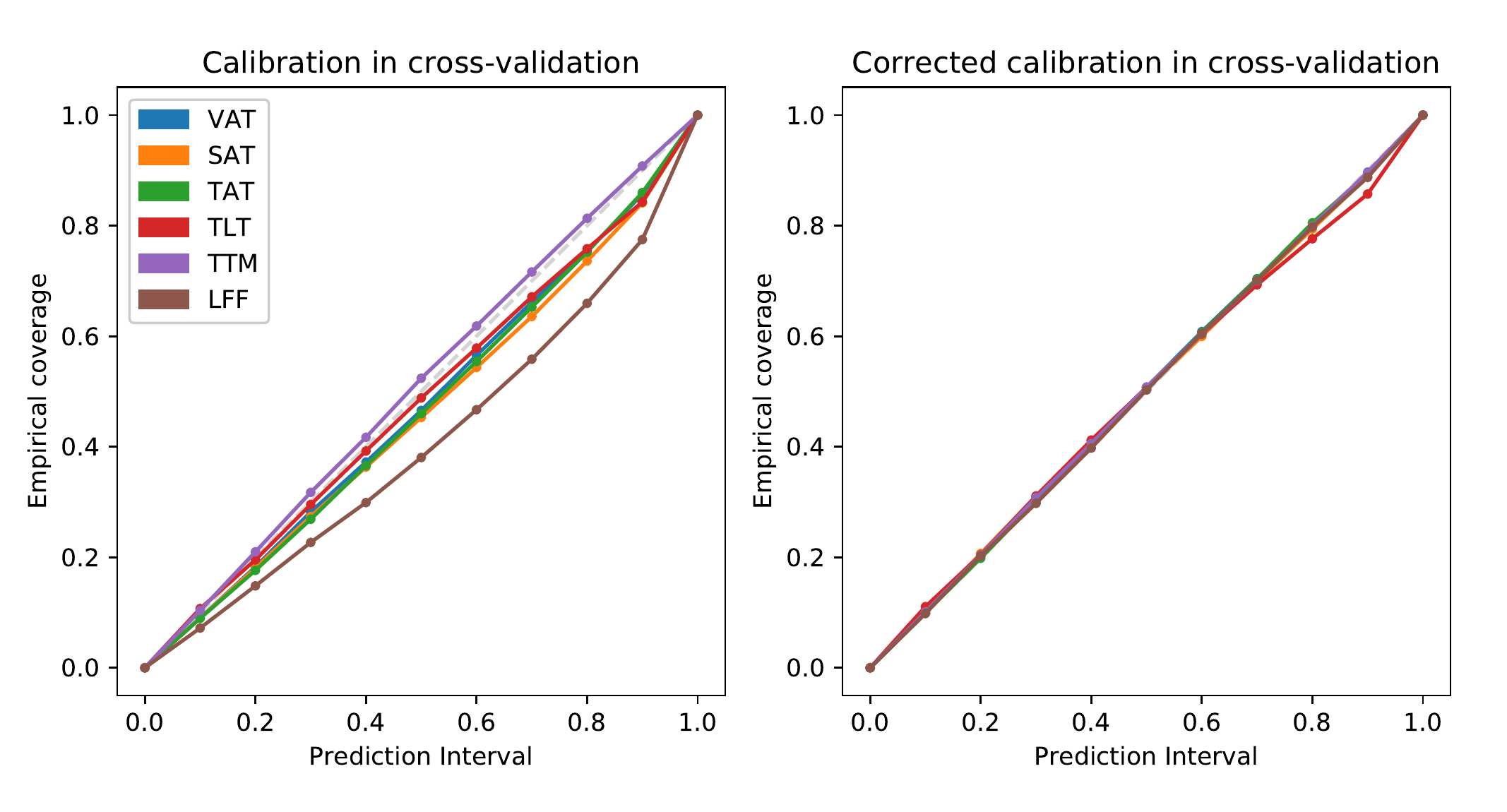}		
\includegraphics[width=0.666\textwidth]{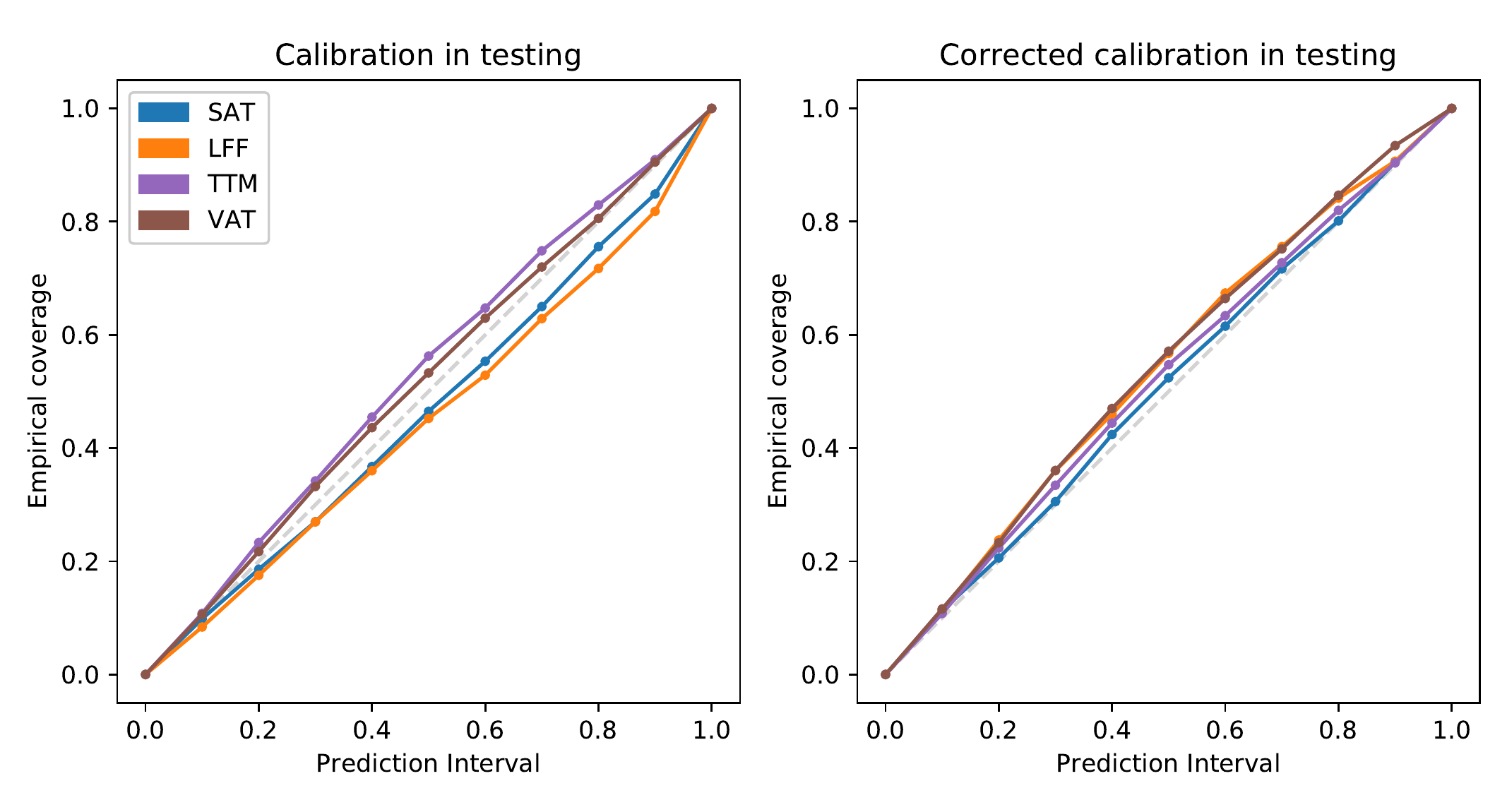}			
\caption{Calibration plots for the \textit{mean-variance} regression ensemble on cross-validation dataset \textit{$D_{cv}$} and testing on dataset \textit{$D_{test}$}. Ideally, each prediction interval as modeled by the underlying predicted Gaussian probability distribution should cover the corresponding share of reference values. This hypothetical optimum is represented by the gray dashed line.}
\label{supp_fig_calibration_fix}	
\end{figure}

\vspace{-0.75cm}

\begin{table}[H]

\begin{center}

\renewcommand{\arraystretch}{1}

\caption{Calibration}

\label{supp_tab_calibration}

\begin{tabular}{lc|cc|cc|c}

\hline

\hline

& & \multicolumn{2}{c}{Cross-validation} & \multicolumn{2}{c}{Testing} & \\

Target & & AUCE & AUCE$_{scaled}$ & AUCE & AUCE$_{scaled}$ & Scaling factor \\

\hline			

Visceral Adipose Tissue & (VAT) & 0.025 & 0.004 & 0.017 & 0.041 & 1.212\\

Abdominal Subcutaneous Adipose Tissue & (SAT) & 0.035 & 0.005 & 0.028 & 0.010 & 1.306\\				

Total Adipose Tissue & (TAT) & 0.029 & 0.003 & &  & 1.250\\			

Total Lean Tissue & (TLT) & 0.017 & 0.011 &&& 1.109\\

Total Thigh Muscle & (TTM) & 0.012 & 0.003 & 0.030 & 0.022 & 0.941\\

Liver Fat Fraction & (LFF) & 0.083 & 0.003 & 0.042 & 0.038 & 1.828\\

\hline

\hline

\multicolumn{7}{l}{* Calibration of the \textit{mean-variance} regression ensemble in cross-validation on dataset \textit{$D_{cv}$} and testing on  dataset \textit{$D_{test}$}.} \\

\multicolumn{7}{l}{The area under calibration error curve (AUCE) can be far reduced (to AUCE$_{scaled}$) with target-wise scaling factors.} \\

\end{tabular}

\end{center}

\end{table}

\textbf{Note:} Earlier versions of this manuscript reported slightly worse calibration metrics due to flawed reversing of the standard scaling.

\subsection{Ensemble Size}

\begin{figure}[H]

\centering	

\includegraphics[width=0.666\textwidth]{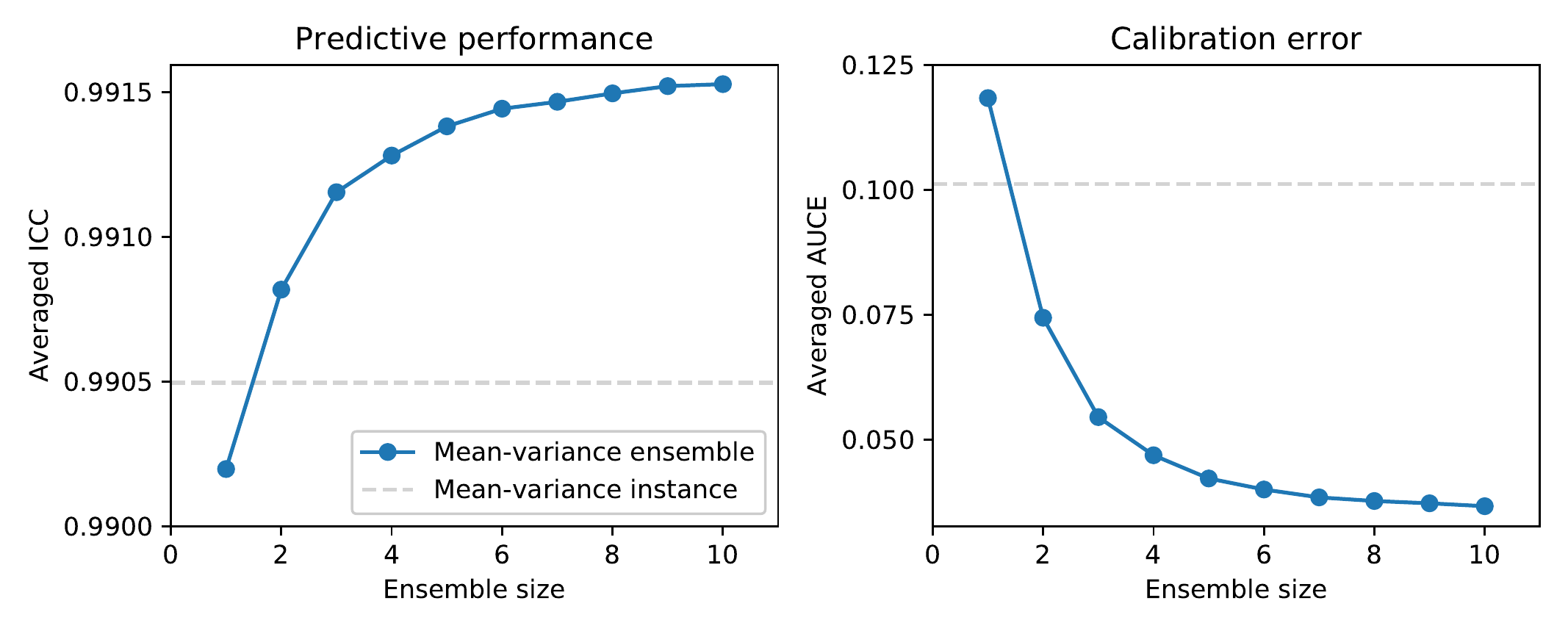}	

\caption{Ablation experiments show how ensembling of mean-variance network instances (blue), each trained on 90\% of samples, compares to a single instance trained on all samples (dotted gray). Averaged across all targets, even ensembles of size two reach superior agreement and calibration.}	

\label{supp_fig_ens_size}

\end{figure}

\subsection{Correlations}

\begin{figure}[h]

\centering	

\includegraphics[width=0.666\textwidth]{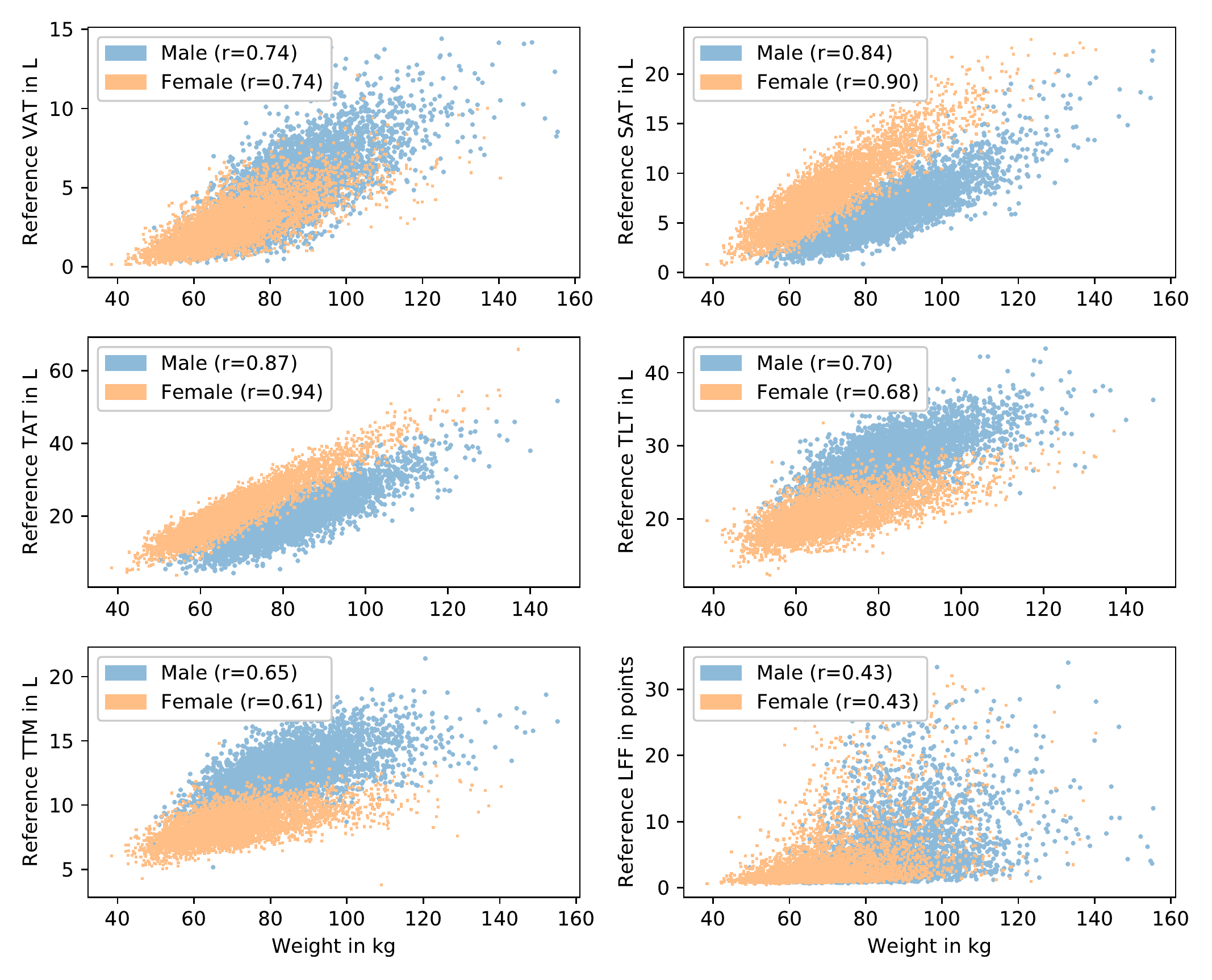}	

\caption{The male- or female-specific target values are highly correlated with body weight (UK Biobank field 21002), as indicated by Pearson's correlation coefficient $r$.}	

\label{supp_fig_cor}

\end{figure}

\newpage

\subsection{Detail on Individual Targets}

The following pages list dedicated plots for the prediction, sparsification, and calibration of each target. For the test data only the \textit{mean-variance} ensemble configuration is shown, which was determined to be the best performing approach in cross-validation.

Each subsection also includes short discussions and comparisons to alternative reference measurements which are primarily derived from two main sources. The first source contains body composition measurements obtained by Dual-energy X-ray absorptiometry (DXA) as conducted by UK Biobank \citep{littlejohns2020uk}. The second source contains additional measurements based on independent machine learning analysis of the same neck-to-knee body MRI as used in this work as conducted by Application 23889, who have shared a return dataset 981.

\hspace{1em}Similar comparisons have been previously reported for a comparable \textit{least squares} regression technique \citep{Langner2020}. Some measurements may be highly correlated but yield low agreement due to a shift or scaling difference. Where specified, these alternative measurements were therefore mapped with linear regression to the target values as used in this work, so that agreement values can be reported. Additionally, Pearson's coefficient of correlation r is reported. For a fair comparison, the methods are evaluated on the same subjects.

\hspace{1em}The sparsification plots also show \textit{oracle sparsification curves}  \citep{ilg2018uncertainty}, which describe a hypothetical optimum that would result from sparsifying with a ranking of uncertainties that corresponds exactly to a ranking of absolute prediction errors. This optimum can typically not be reached in practice, as it would require imitating not only the desired measurements but also any inconsistencies and noise in the reference techniques themselves. The sparsification for the three evaluation datasets is shown separately, but it is worth noting that in most cases the samples with artifacts incurred the highest uncertainty. When applied to a dataset that included mixed normal material and artifacts, the latter would therefore typically be excluded first in the sparsification. The outlier with largest prediction error in testing for VAT, SAT, and TAT is the same subject, found to suffer from an atrophied right leg.

Aggregated saliency maps were obtained by generating guided gradient-weighted class activation maps for 3,091 subjects and co-aligning them by image registration  \citep{Langner2019}. Each aggregated saliency map accordingly highlights which anatomical structures were predominantly considered by the network to make predictions for the specified target. For clarity, the visualizations show the aggregated saliency as a heatmap for each of the three input image channels side by side and are provided with and without the template subject anatomy as an overlay. The network weights used for this purpose are based on the \textit{mean-variance} configuration with a single network trained for cross-validation in this work, in each case using the instance that did not contain the given image in its training set.

\newpage

\setlength{\abovecaptionskip}{-0.1cm}

%%%%%%%%%%%%%%%%%%%%%%%%%%%%%%%%%%%%%%%%%%%%%%%%%%%%%%%%%%%%%%%%%%%%%%%%%%%%%%%%%%%%%%%%%%%%%%%

% VAT

\newpage

\subsubsection{\Large \textbf{Visceral Adipose Tissue (VAT)}}

\begin{figure}[H]
\centering	
\includegraphics[width=\textwidth]{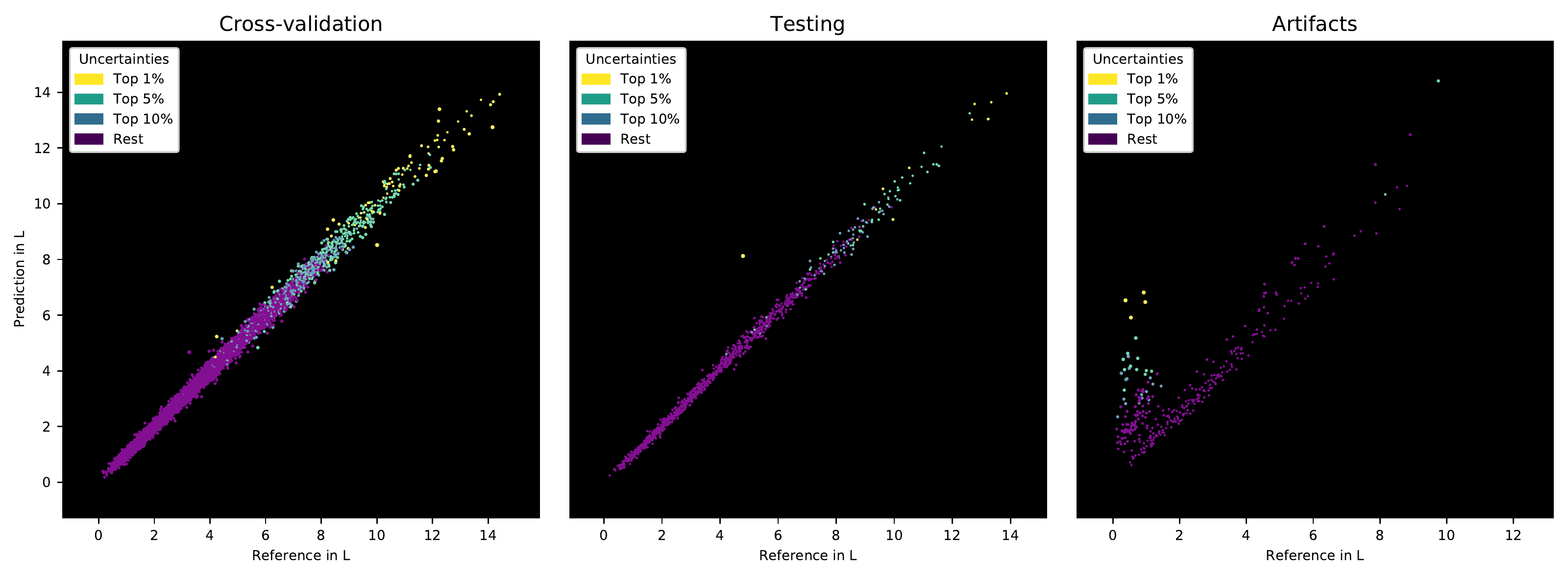}		
\caption{Predictions in cross-validation on \textit{$D_{cv}$}, testing on \textit{$D_{test}$}, and on subjects with artifacts of \textit{$D_{art}$}, with color-coded uncertainty. }
\label{supp_fig_scatter_vat}	
\end{figure}

\begin{figure}[H]
\centering	
\includegraphics[width=\textwidth]{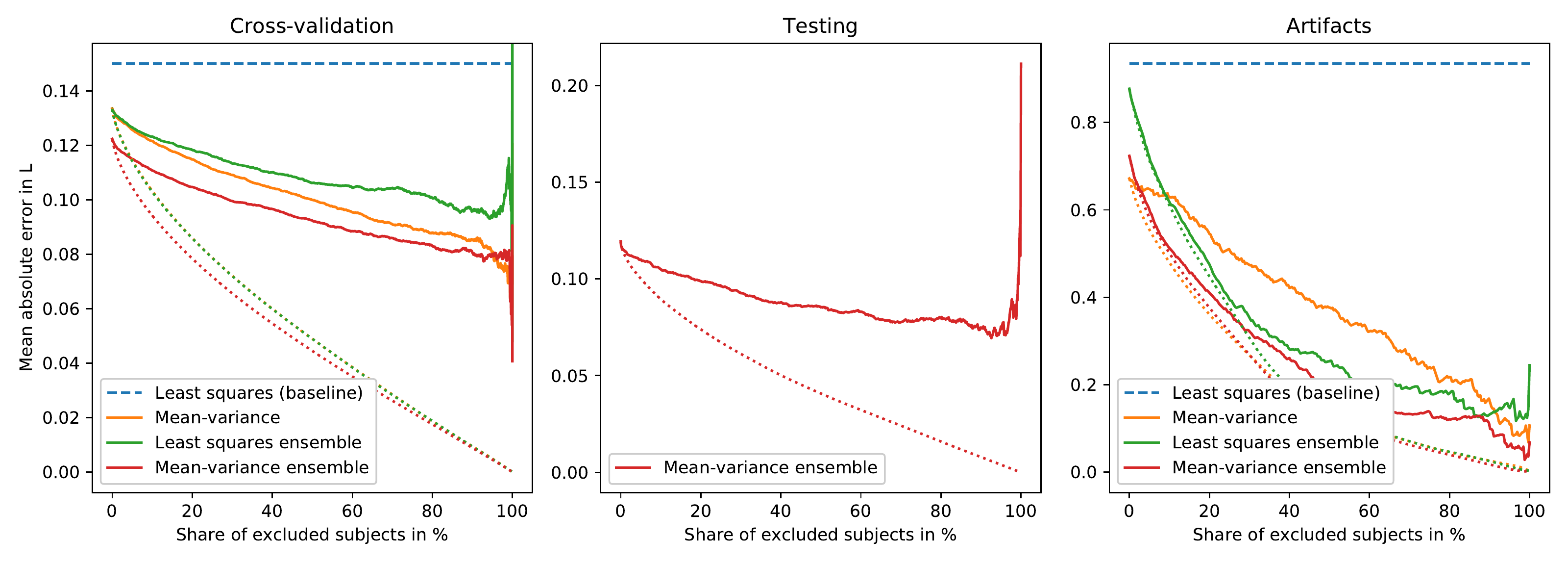}		
\caption{Sparsification in cross-validation on \textit{$D_{cv}$}, testing on \textit{$D_{test}$}, and on subjects with artifacts of \textit{$D_{art}$}, with oracle curves (dotted).}
\label{supp_fig_sparsify_vat}	
\end{figure}

\begin{figure}[H]
\centering	
\includegraphics[width=\textwidth]{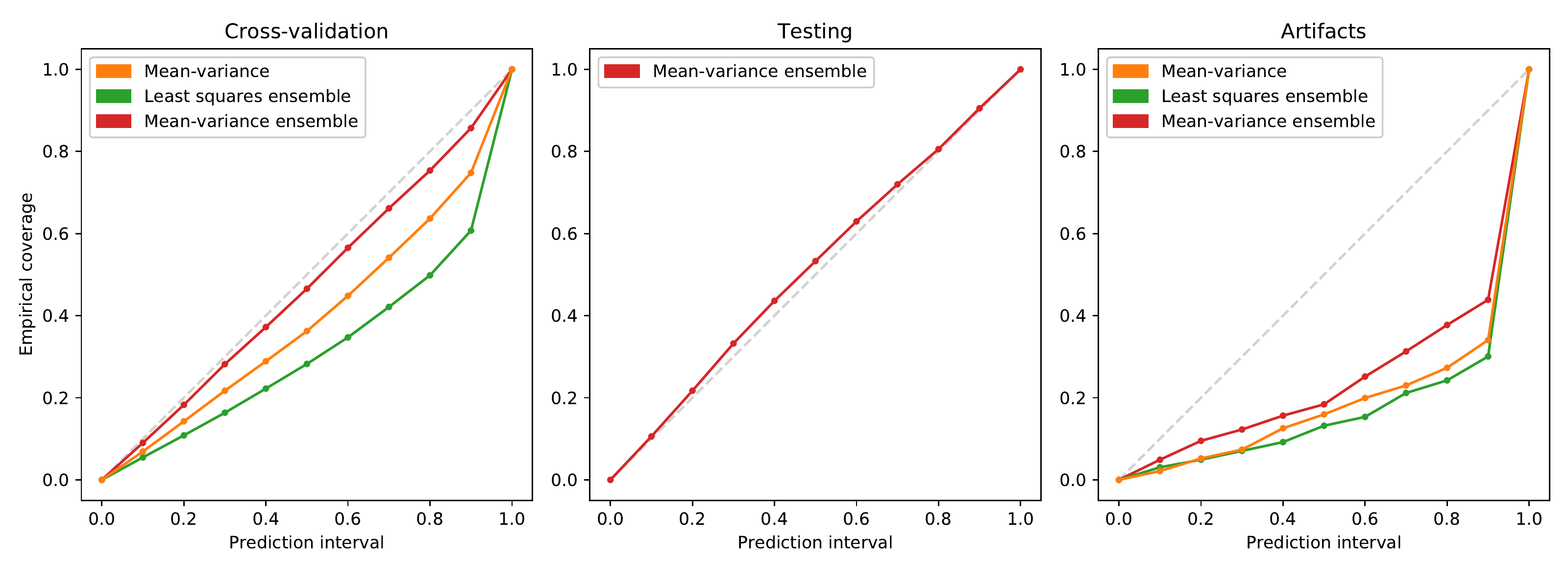}		
\caption{Calibration in cross-validation on \textit{$D_{cv}$}, testing on \textit{$D_{test}$}, and on subjects with artifacts of \textit{$D_{art}$}.}
\label{supp_fig_calibration_vat}	
\end{figure}

\textbf{Visceral Adipose Tissue (VAT)}, extended notes:

Supplementary Fig. \ref{supp_fig_scatter_vat} shows a close fit with few outliers in the normal material. In testing, a single subject with an atrophied right leg incurs a substantially overestimated measurement, which can be identified by high uncertainty. \\

\textbf{Alternative reference methods:}

UK Biobank field 23289 contains measurements of VAT by DXA for 5,109 subjects. These values were first converted from mL to L and then mapped to the target with the following linear transformation parameters:

($2.27x + 0.83 L$).

UK Biobank return 981 by application 23889 also offers VAT measurements for 9,127 subjects. These values were converted from mL to L, but did not require adjustment by linear regression.

\begin{table}[H]

\begin{center}	

\renewcommand{\arraystretch}{1.25}	

\caption{Comparison of VAT references}	

\label{tab_supp_alt_vat}

\begin{tabular}{llrrrrr}

\hline	

\hline	

Method & N &  ICC & R2 & MAE & MAPE & r\\

\hline			

Proposed & 4,491 & \textbf{0.997} & \textbf{0.994} & \textbf{0.131} & \textbf{4.3} & \textbf{0.997}  \\

Field 23289 & 4,491 & 0.970 & 0.942 & 0.401 & 14.9 & 0.971 \\

\hline

Proposed & 

7,871 & \textbf{0.997} & \textbf{0.994} & \textbf{0.121} & \textbf{4.1} & \textbf{0.997} \\

Return 981 & %N,ICC,R2,MAE,MAPE,r

7,871 & 0.996 & 0.993 & 0.137 & 4.4 & 0.996 \\

\hline	

\hline	

\multicolumn{7}{l}{*Comparison to the target values, listing both the proposed predictions}\\  \multicolumn{7}{l}{  and alternative UK Biobank reference values on the same subjects}

\end{tabular}

\end{center}

\end{table}

\textbf{Aggregated saliency (VAT):}

\begin{figure}[H]

\centering	

\includegraphics[width=\textwidth]{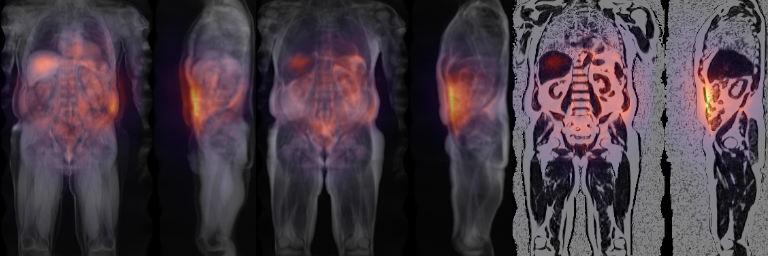}			

\includegraphics[width=\textwidth]{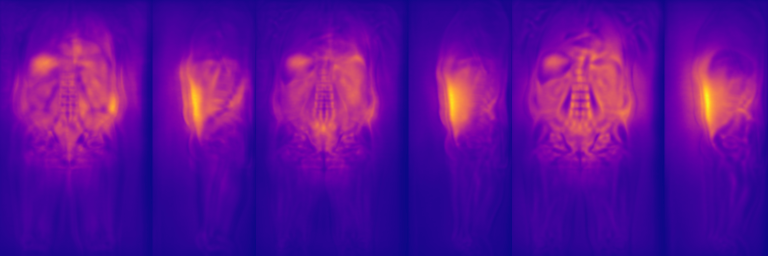}					

\caption{Aggregated saliency \citep{Langner2019} for Visceral Adipose Tissue (VAT) for 3,091 subjects, generated by a single \textit{mean-variance} network. Each row shows the water, fat, and fat-fraction channels side by side, with the top row showing an overlay on the image data and the bottom row the saliency only.}

\label{supp_fig_sal_vat}	

\end{figure}

\newpage

%%%%%%%%%%%%%%%%%%%%%%%%%%%%%%%%%%%%%%%%%%%%%%%%%%%%%%%%%%%%%%%%%%%%%%%%%%%%%%%%%%%%%%%%%%%%%%%

% SAT

\subsubsection{\Large \textbf{Abdominal Subcutaneous Adipose Tissue (SAT)}}

\begin{figure}[H]

\centering	

\includegraphics[width=\textwidth]{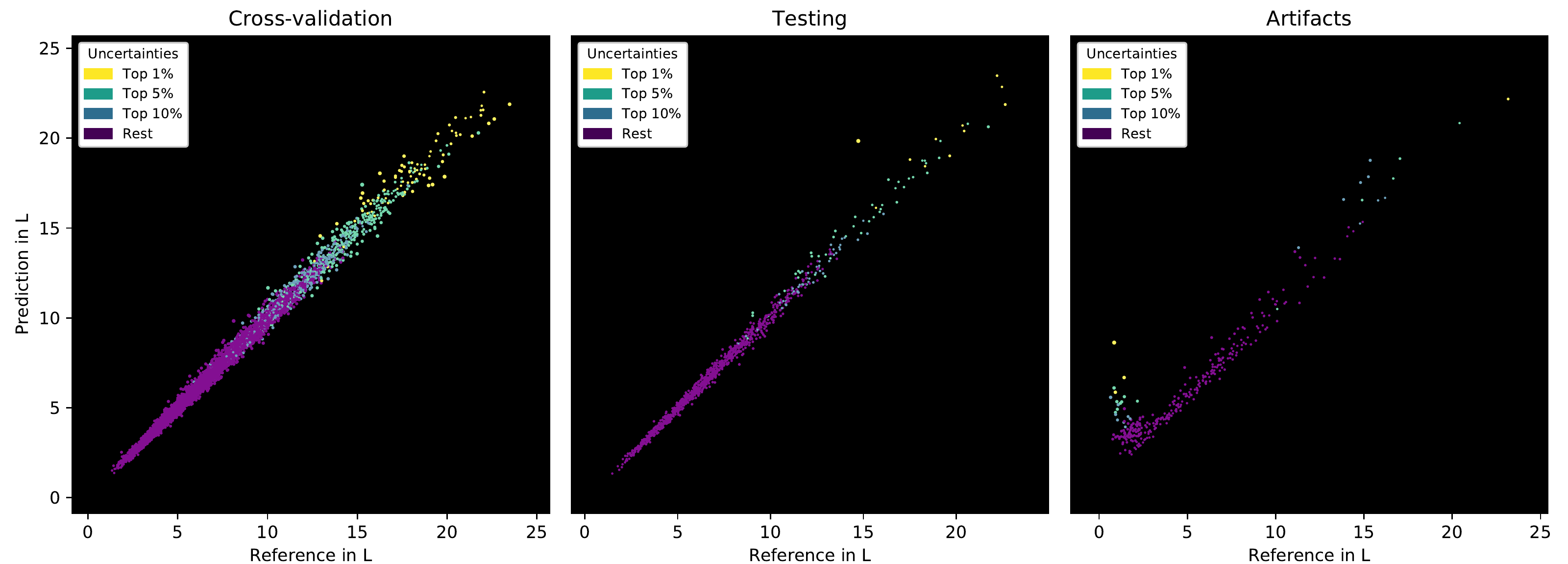}		

\caption{Predictions in cross-validation on \textit{$D_{cv}$}, testing on \textit{$D_{test}$}, and on subjects with artifacts of \textit{$D_{art}$}, with color-coded uncertainty. }

\label{supp_fig_scatter_sat}	

\end{figure}

\begin{figure}[H]

\centering	

\includegraphics[width=\textwidth]{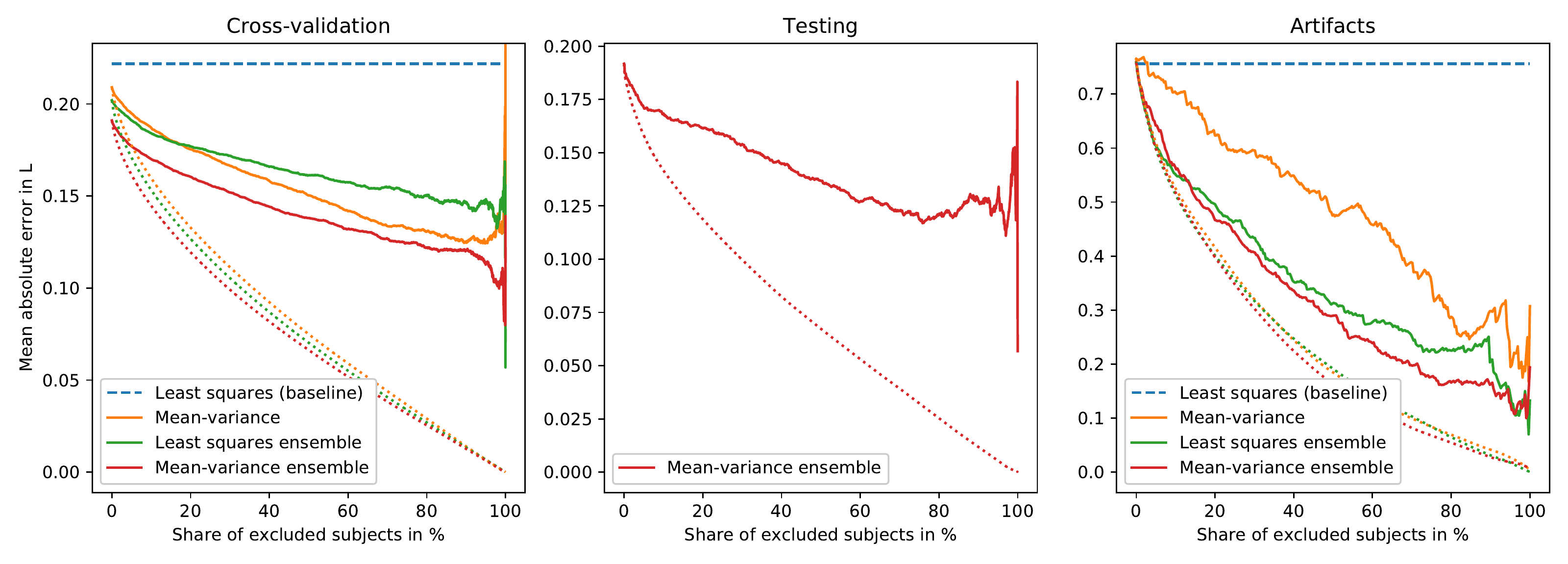}		

\caption{Sparsification in cross-validation on \textit{$D_{cv}$}, testing on \textit{$D_{test}$}, and on subjects with artifacts of \textit{$D_{art}$}, with oracle curves (dotted).}

\label{supp_fig_sparsify_sat}	

\end{figure}

\begin{figure}[H]

\centering	

\includegraphics[width=\textwidth]{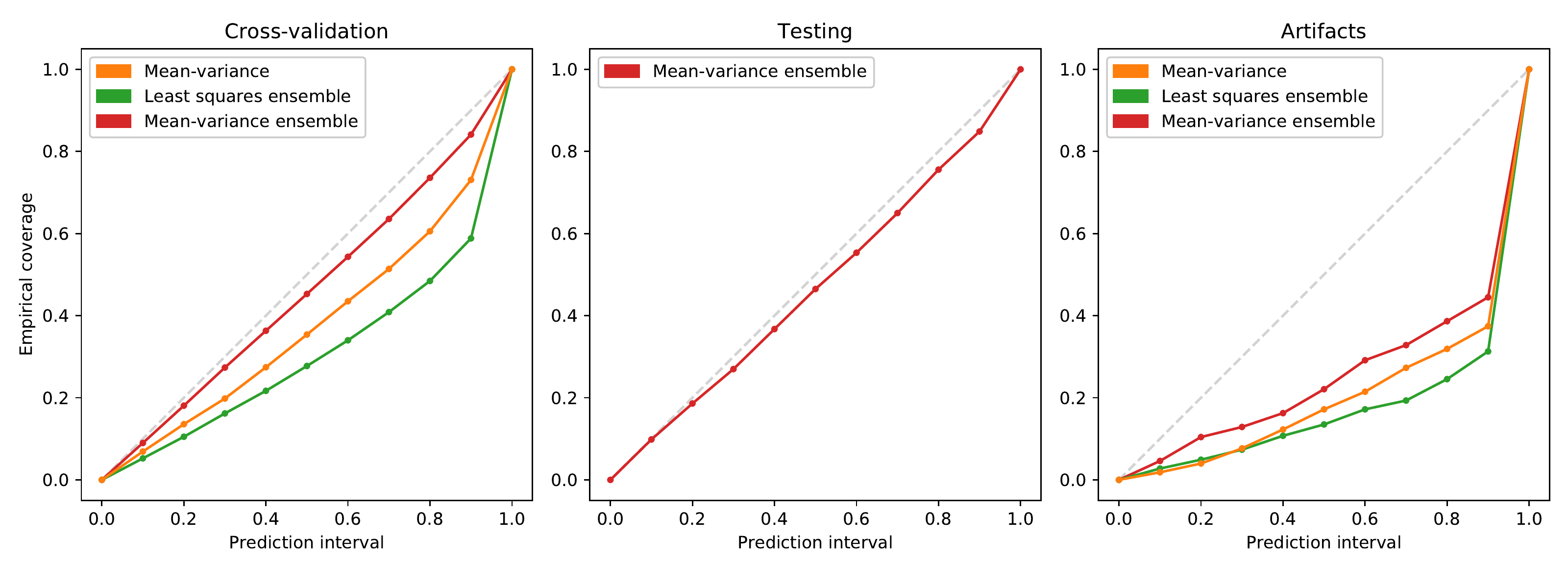}		

\caption{Calibration in cross-validation on \textit{$D_{cv}$}, testing on \textit{$D_{test}$}, and on subjects with artifacts of \textit{$D_{art}$}.}

\label{supp_fig_calibration_sat}	

\end{figure}

\newpage

\textbf{Abdominal Subcutaneous Adipose Tissue (SAT)}, extended notes:

The scatter plot for the test data of Fig.\ref{supp_fig_scatter_sat} shows a single outlier with about 15 L of subcutaneous adipose tissue, for whom the prediction yields almost 20 L with high uncertainty. This subject was found to suffer from an abnormal, atrophied right leg and also incurs high measurement errors in TTM and VAT. \\

\textbf{Alternative reference methods:}

UK Biobank return 981 by application 23889 also offers measurements of subcutaneous adipose tissue volume for 9,379 subjects. These values were converted from mL to L and then mapped to the target with the following linear transformation parameters: ($0.98x + 0.46 L$).

\begin{table}[H]

\begin{center}	

\renewcommand{\arraystretch}{1.25}

\caption{Comparison of SAT references}

\label{tab_supp_alt_sat}

\begin{tabular}{llrrrrr}

\hline	

\hline	

Method & N &  ICC & R2 & MAE & MAPE & r\\

\hline			

Proposed & 8,085 & \textbf{0.996} & \textbf{0.993} & \textbf{0.187} & \textbf{2.8} & \textbf{0.996} \\

Return 981 & 8,085 & 0.994 & 0.989 & 0.208 & 3.1 & 0.994 \\

\hline	

\hline

\multicolumn{7}{l}{*Comparison to the target values, listing both the proposed predictions}\\  \multicolumn{7}{l}{  and alternative UK Biobank reference values on the same subjects}

\end{tabular}

\end{center}

\end{table}

\textbf{Aggregated saliency (SAT):}

\begin{figure}[H]

\centering	

\includegraphics[width=\textwidth]{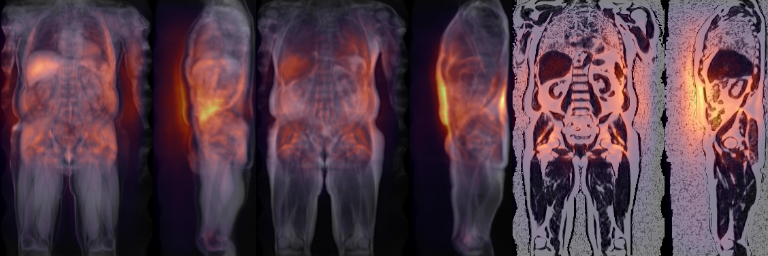}			

\includegraphics[width=\textwidth]{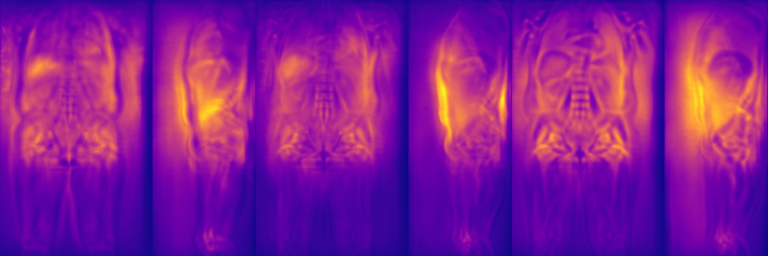}					

\caption{Aggregated saliency \citep{Langner2019} for Subcutaneous Adipose Tissue (SAT) for 3,091 subjects, generated by a single \textit{mean-variance} network. Each row shows the water, fat, and fat-fraction channels side by side, with the top row showing an overlay on the image data and the bottom row the saliency only.}

\label{supp_fig_sal_sat}	

\end{figure}

\iffalse

Klarismo vs Amra

N,ICC,R2,MAE,MAPE,r

8085,0.9874360293173091,0.9745300374877416,0.40780729746444033,0.0673411820350503,0.9943927374603992

Own vs Amra

N,ICC,R2,MAE,MAPE,r

8085,0.9963791017759239,0.9927869360340886,0.1869748473969056,0.027750405136482543,0.9963910081625839

\fi

%%%%%%%%%%%%%%%%%%%%%%%%%%%%%%%%%%%%%%%%%%%%%%%%%%%%%%%%%%%%%%%%%%%%%%%%%%%%%%%%%%%%%%%%%%%%%%%

% TAT

\newpage

\subsubsection{\Large \textbf{Total Adipose Tissue (TAT)}}

\begin{figure}[H]

\centering	

\includegraphics[width=\textwidth]{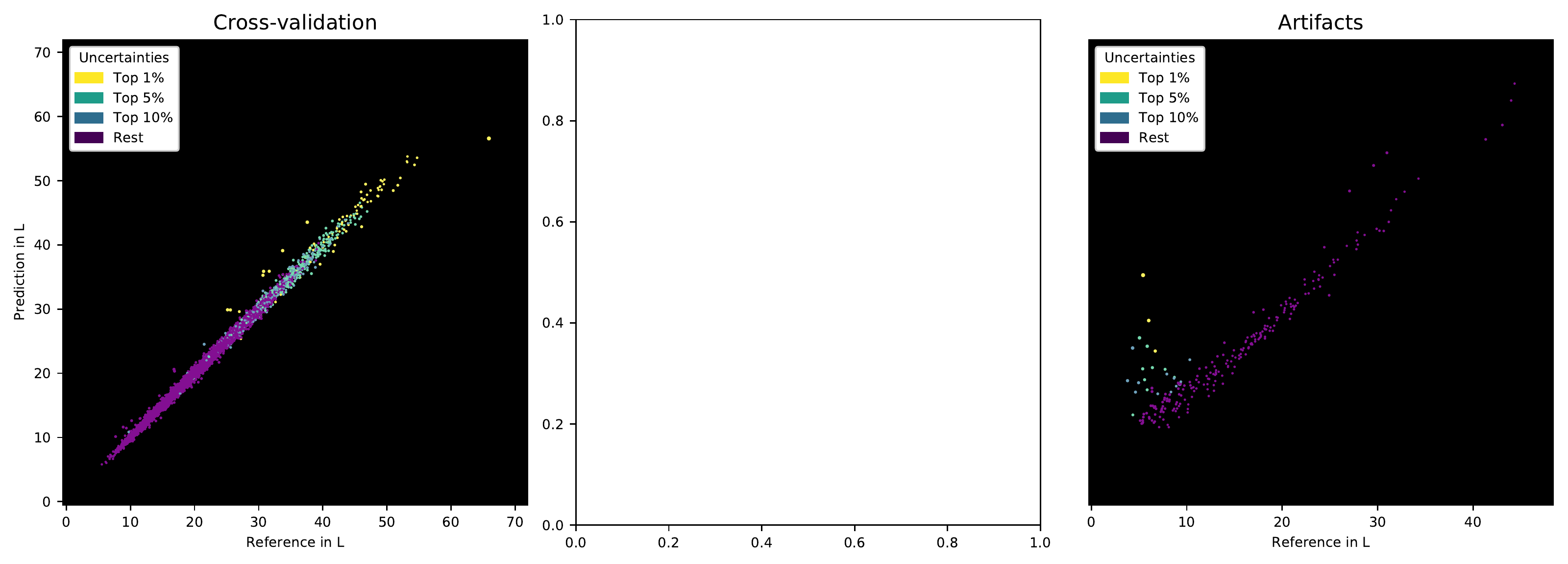}		

\caption{Predictions in cross-validation on \textit{$D_{cv}$}, testing on \textit{$D_{test}$}, and on subjects with artifacts of \textit{$D_{art}$}, with color-coded uncertainty. }

\label{supp_fig_scatter_tat}	

\end{figure}

\begin{figure}[H]

\centering	

\includegraphics[width=\textwidth]{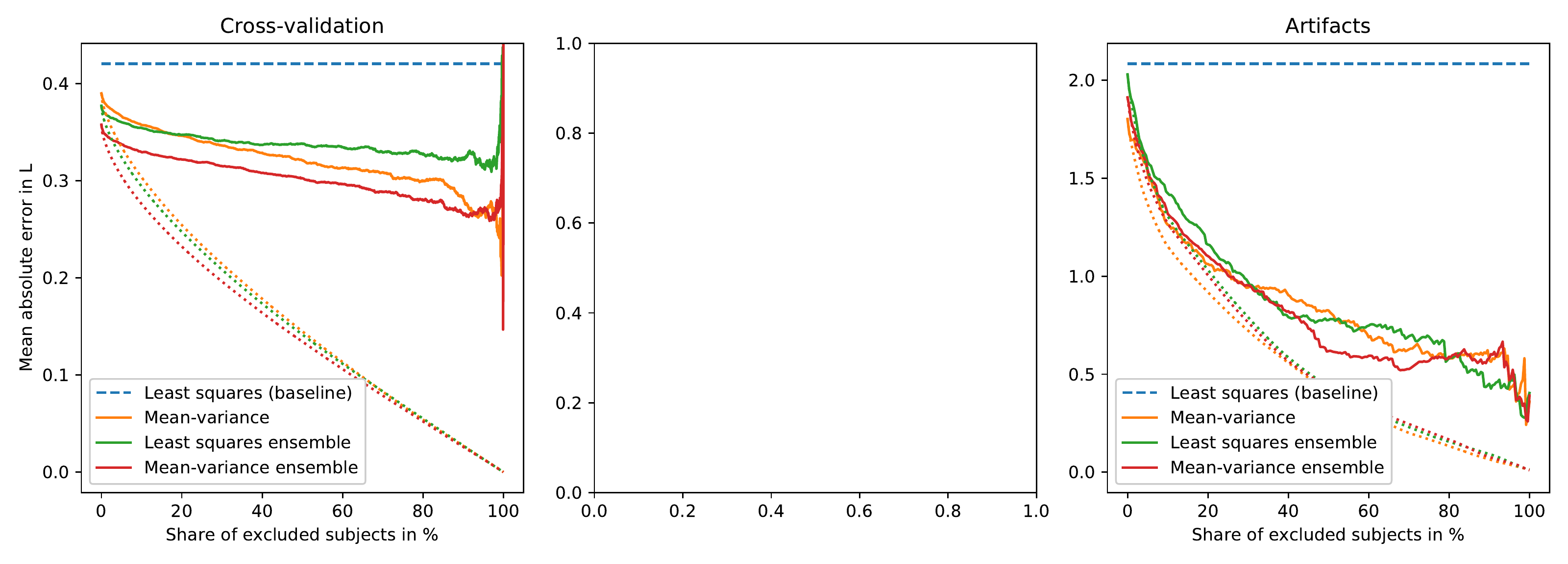}		

\caption{Sparsification in cross-validation on \textit{$D_{cv}$}, testing on \textit{$D_{test}$}, and on subjects with artifacts of \textit{$D_{art}$}, with oracle curves (dotted).}

\label{supp_fig_sparsify_tat}	

\end{figure}

\begin{figure}[H]

\centering	

\includegraphics[width=\textwidth]{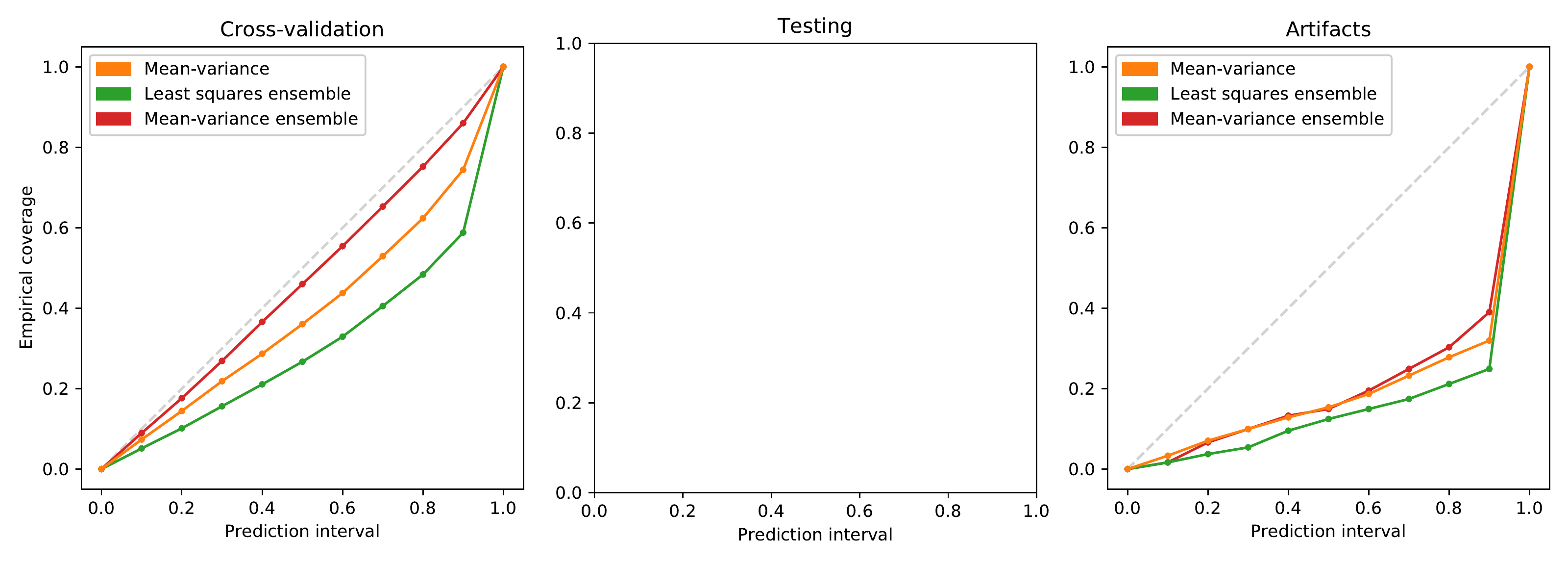}		

\caption{Calibration in cross-validation on \textit{$D_{cv}$}, testing on \textit{$D_{test}$}, and on subjects with artifacts of \textit{$D_{art}$}.}

\label{supp_fig_calibration_tat}	

\end{figure}

\pagebreak

\textbf{Total Adipose Tissue (TAT)}, extended notes:

No test data was available for this target. \\

\textbf{Alternative reference methods:}

UK Biobank field 23278 contains alternative measurements of total fat mass by DXA for 5,170 subjects. These values were first converted from mL to L and then mapped to the target with the following linear transformation parameters:

($0.80x + 0.51 L$).

\begin{table}[H]

\begin{center}	

\renewcommand{\arraystretch}{1.25}

\caption{Comparison of TLT references}

\label{tab_supp_alt_tat}

\begin{tabular}{llrrrrr}

\hline	

\hline	

Method & N &  ICC & R2 & MAE & MAPE & r\\

\hline			

Proposed & 4,323 & \textbf{0.997} & \textbf{0.995} & \textbf{0.353} & \textbf{1.8} & \textbf{0.997} \\

Field 23278 & 4,323 & 0.991 & 0.982 & 0.689 & 3.4 & 0.991 \\

\hline	

\hline	

\multicolumn{7}{l}{*Comparison to the target values, listing both the proposed predictions}\\  \multicolumn{7}{l}{  and alternative UK Biobank reference values on the same subjects}	

\end{tabular}

\end{center}

\end{table}

\textbf{Aggregated saliency (TAT):}

\begin{figure}[H]

\centering	

\includegraphics[width=\textwidth]{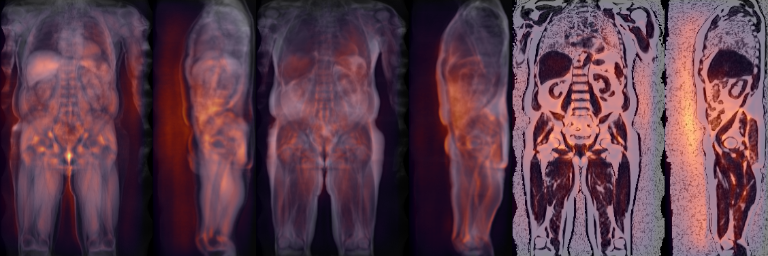}			

\includegraphics[width=\textwidth]{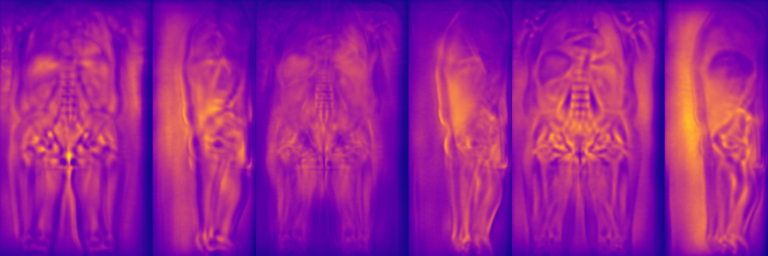}					

\caption{Aggregated saliency \citep{Langner2019} for Total Adipose Tissue (TAT) for 3,091 subjects, generated by a single \textit{mean-variance} network. Each row shows the water, fat, and fat-fraction channels side by side, with the top row showing an overlay on the image data and the bottom row the saliency only.}

\label{supp_fig_sal_tat}	

\end{figure}

\newpage

%%%%%%%%%%%%%%%%%%%%%%%%%%%%%%%%%%%%%%%%%%%%%%%%%%%%%%%%%%%%%%%%%%%%%%%%%%%%%%%%%%%%%%%%%%%%%%%

% TLT

\newpage

\subsubsection{\Large \textbf{Total Lean Tissue (TLT)}}

\begin{figure}[H]

\centering	

\includegraphics[width=\textwidth]{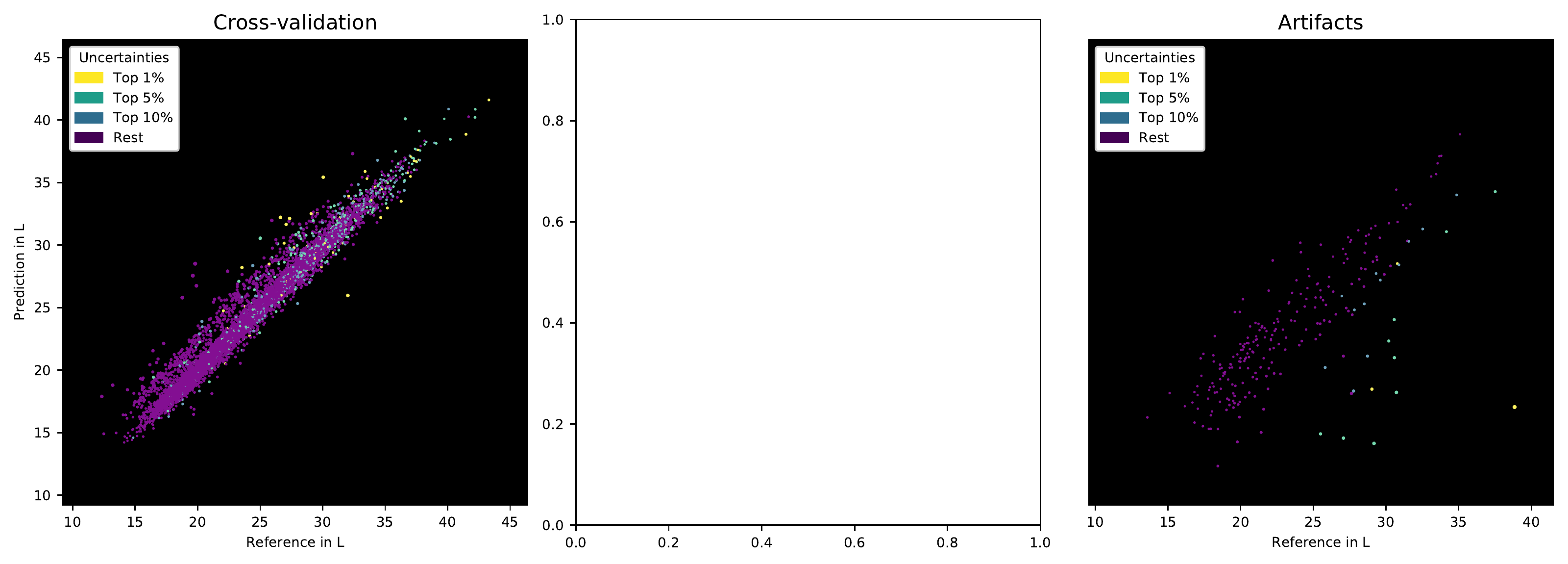}		

\caption{Predictions in cross-validation on \textit{$D_{cv}$}, testing on \textit{$D_{test}$}, and on subjects with artifacts of \textit{$D_{art}$}, with color-coded uncertainty. }

\label{supp_fig_scatter_tlt}	

\end{figure}

\begin{figure}[H]

\centering	

\includegraphics[width=\textwidth]{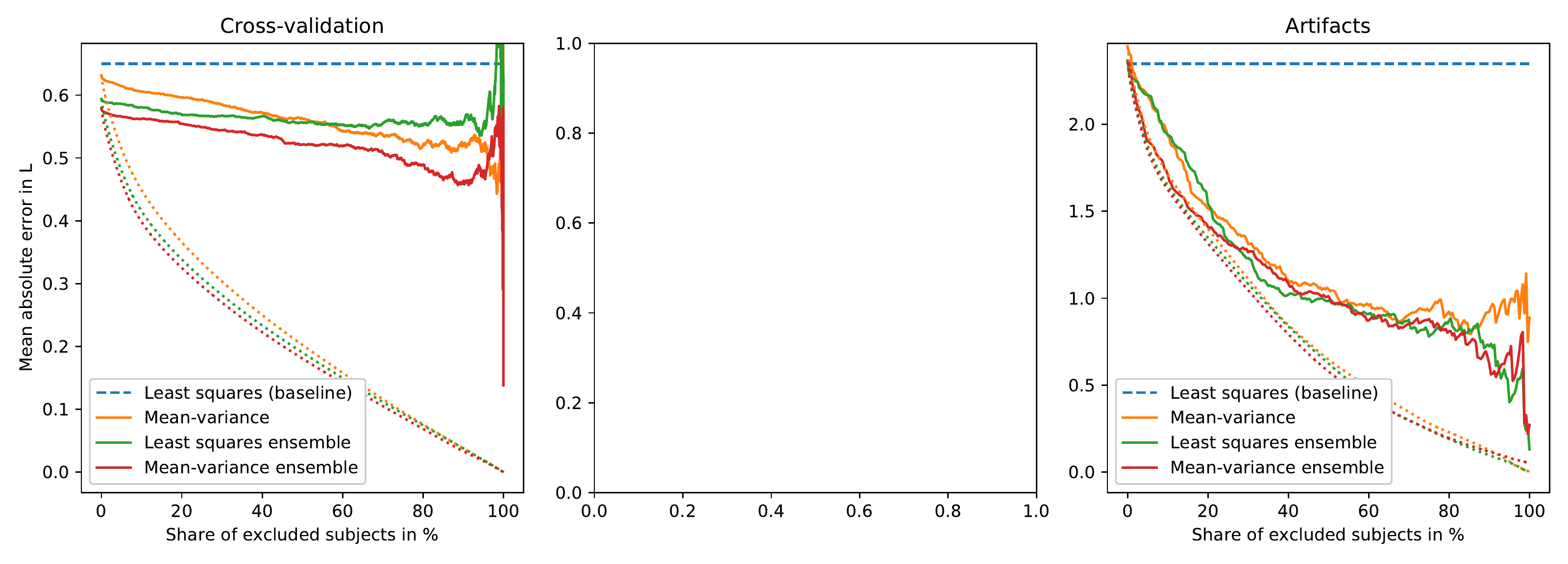}		

\caption{Sparsification in cross-validation on \textit{$D_{cv}$}, testing on \textit{$D_{test}$}, and on subjects with artifacts of \textit{$D_{art}$}, with oracle curves (dotted).}

\label{supp_fig_sparsify_tlt}	

\end{figure}

\begin{figure}[H]

\centering	

\includegraphics[width=\textwidth]{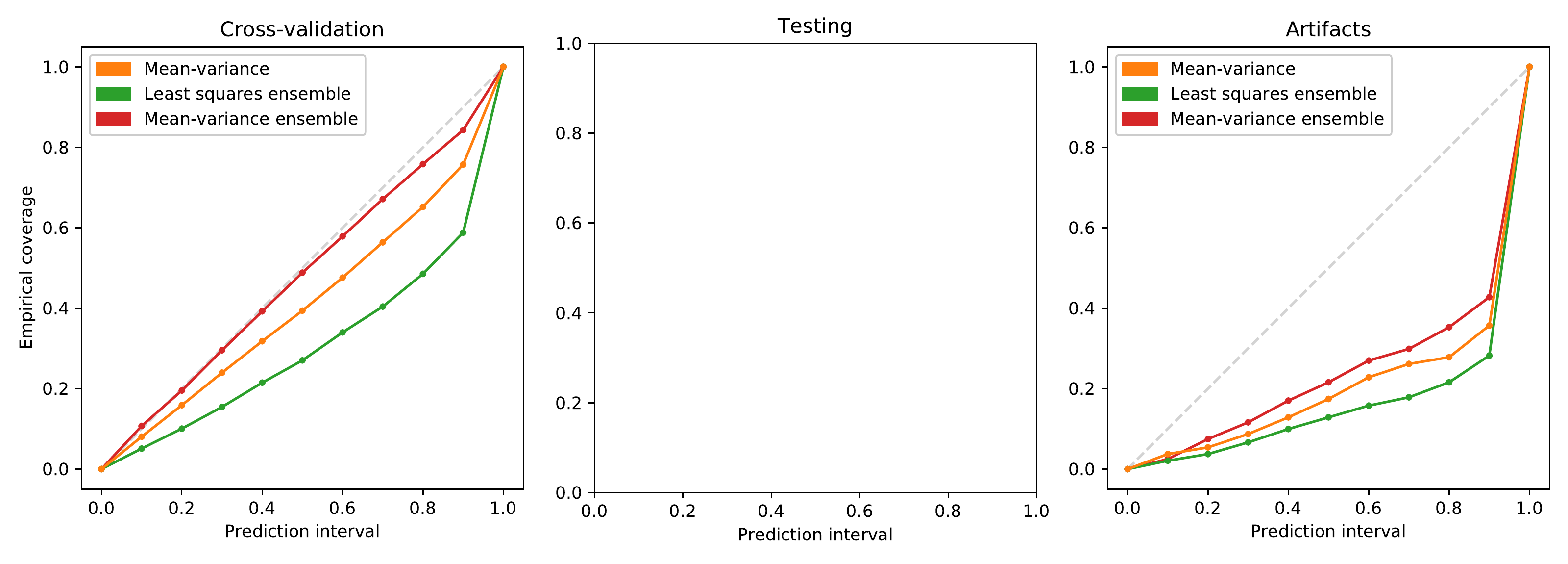}		

\caption{Calibration in cross-validation on \textit{$D_{cv}$}, testing on \textit{$D_{test}$}, and on subjects with artifacts of \textit{$D_{art}$}.}

\label{supp_fig_calibration_tlt}	

\end{figure}

\pagebreak

\textbf{Total Lean Tissue (TLT)}, extended notes:

No test data was available for this target. Supplementary Fig. \ref{supp_fig_scatter_tlt} shows a curious pattern for the cross-validation, where a subset of measurements is consistently overestimated by about 2 L.

\begin{figure}[H]

\centering	

\includegraphics[width=0.66\textwidth]{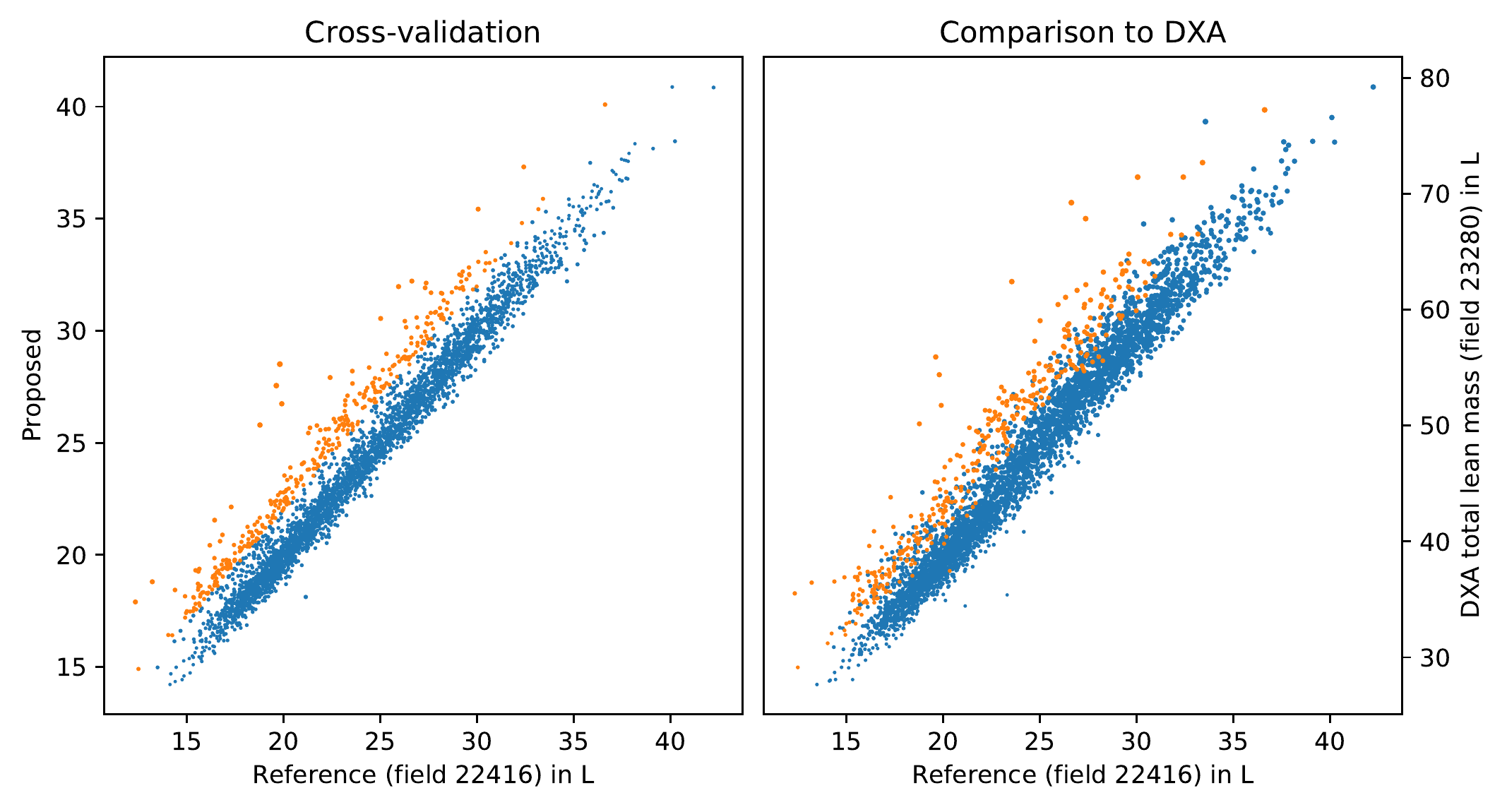}		

\caption{In some subjects (red), the proposed method overestimated total lean tissue (TLT) by at least 2L. As shown on the right, the DXA scan shows a similar pattern and independently indicates higher values for these subjects.}

\label{supp_fig_scatter_tlt_dxa}	

\end{figure}

The reason for this mismatch is unclear. The affected subjects are not part of the same cross-validation split set, were imaged in different imaging centers, and share no other obvious confounding factors. However, alternative measurements of total lean tissue by DXA (total lean mass, field 23280) independently support these overestimations relative to the reference used in this work. Supplementary Fig. \ref{supp_fig_scatter_tlt_dxa} shows a comparison where the reference is plotted against the DXA measurements. All those cases that were overestimated by the proposed method by at least 2L are color-coded and form a similar pattern as observed in cross-validation.

\textbf{Alternative reference methods:}

UK Biobank field 23280 contains additional measurements of total lean mass by DXA for 5,170 subjects. These values were first converted from mL to L and then mapped to the target with the following linear transformation parameters:

($0.50x + 0.47 L$).

On a side note, UK Biobank field 23285 also contains DXA measurements of trunk lean mass, but these values reaches lower agreement with the target than field 23280 and were not considered further.

\begin{table}[H]

\begin{center}	

\renewcommand{\arraystretch}{1.25}	

\caption{Comparison of TLT references}

\label{tab_supp_alt_tlt}

\begin{tabular}{llrrrrr}

\hline	

\hline	

Method & N &  ICC & R2 & MAE & MAPE & r\\

\hline			

Proposed & 4,323 & \textbf{0.976} & \textbf{0.953} & \textbf{0.684} & \textbf{3.0} & \textbf{0.978}  \\

Field 23280 & 4,323 & 0.969 & 0.941 & 0.856 & 3.7 & 0.970 \\

\hline	

\hline	

\multicolumn{7}{l}{*Comparison to the target values, listing both the proposed predictions}\\  \multicolumn{7}{l}{  and alternative UK Biobank reference values on the same subjects}	

\end{tabular}

\end{center}

\end{table}

\newpage

\textbf{Aggregated saliency (TLT):}

\begin{figure}[H]

\centering	

\includegraphics[width=\textwidth]{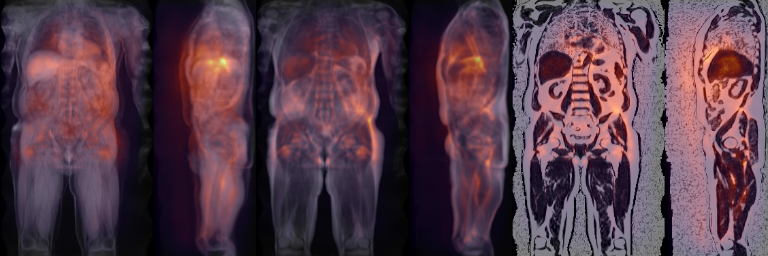}			

\includegraphics[width=\textwidth]{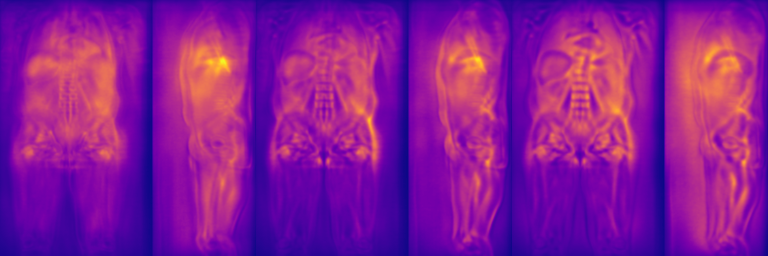}					

\caption{Aggregated saliency \citep{Langner2019} for Total Lean Tissue (TLT) for 3,091 subjects, generated by a single \textit{mean-variance} network. Each row shows the water, fat, and fat-fraction channels side by side, with the top row showing an overlay on the image data and the bottom row the saliency only.}

\label{supp_fig_sal_tlt}	

\end{figure}

\newpage

%%%%%%%%%%%%%%%%%%%%%%%%%%%%%%%%%%%%%%%%%%%%%%%%%%%%%%%%%%%%%%%%%%%%%%%%%%%%%%%%%%%%%%%%%%%%%%%

% TTM

\newpage

\subsubsection{\Large \textbf{Total Thigh Muscle (TTM)}}

\begin{figure}[H]

\centering	

\includegraphics[width=\textwidth]{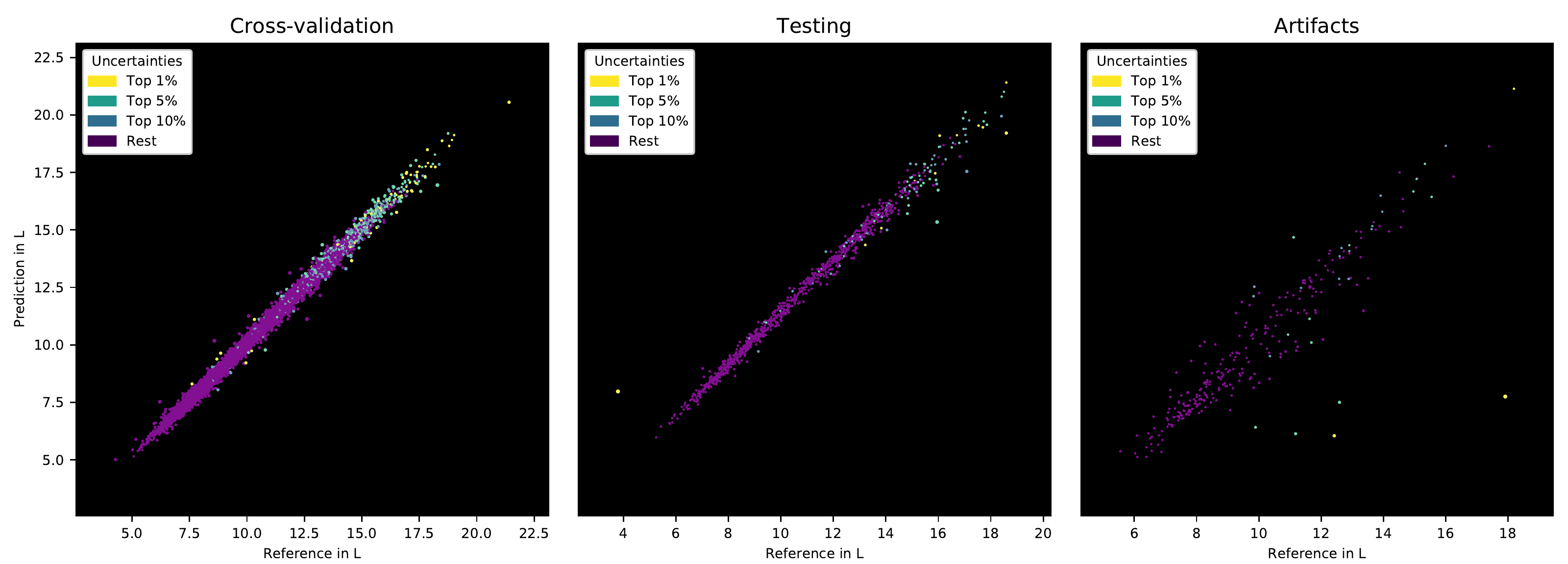}		

\caption{Predictions in cross-validation on \textit{$D_{cv}$}, testing on \textit{$D_{test}$}, and on subjects with artifacts of \textit{$D_{art}$}, with color-coded uncertainty. }

\label{supp_fig_scatter_ttm}	

\end{figure}

\begin{figure}[H]

\centering	

\includegraphics[width=\textwidth]{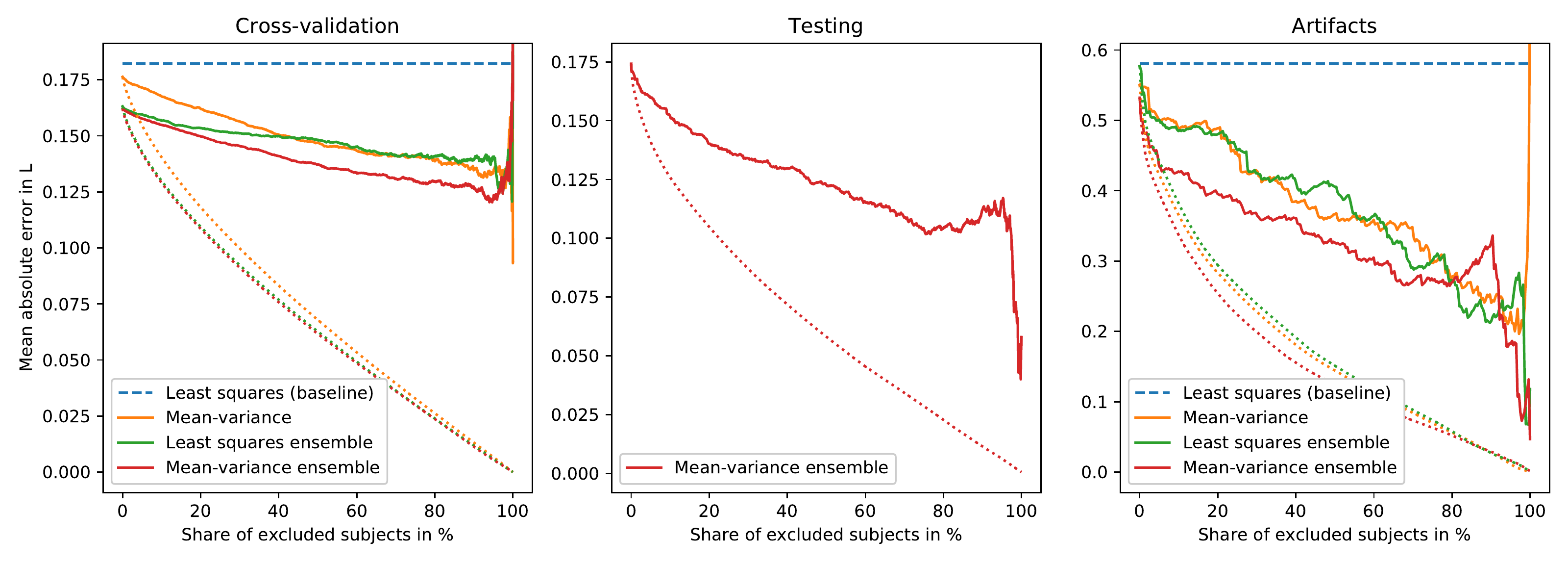}		

\caption{Sparsification in cross-validation on \textit{$D_{cv}$}, testing on \textit{$D_{test}$}, and on subjects with artifacts of \textit{$D_{art}$}, with oracle curves (dotted).}

\label{supp_fig_sparsify_ttm}	

\end{figure}

\begin{figure}[H]

\centering	

\includegraphics[width=\textwidth]{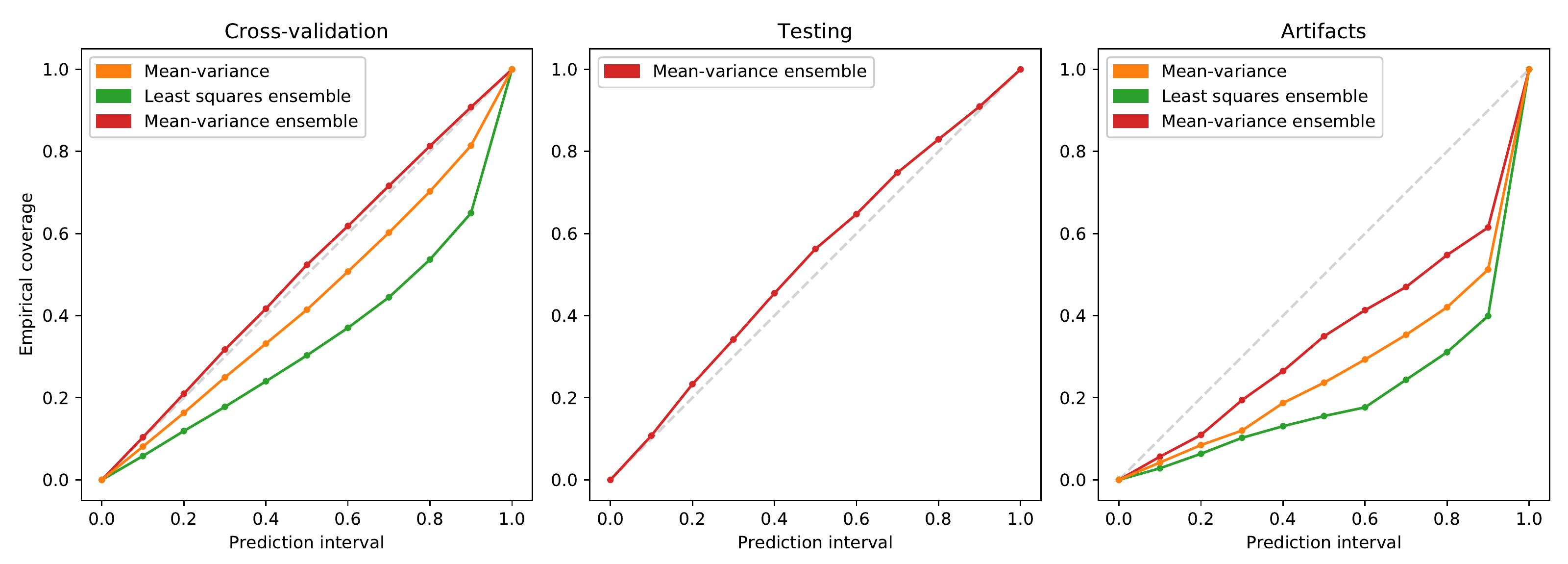}		

\caption{Calibration in cross-validation on \textit{$D_{cv}$}, testing on \textit{$D_{test}$}, and on subjects with artifacts of \textit{$D_{art}$}.}

\label{supp_fig_calibration_ttm}	

\end{figure}

\pagebreak

\textbf{Total Thigh Muscle (TTM)}, extended notes:

Supplementary Fig. \ref{supp_fig_scatter_ttm} shows a close fit with few outliers in the normal material. In testing, a single subject with an atrophied right leg incurs high uncertainty, together with a moderately overestimated measurement. Several other high-valued testing cases are slightly underestimated. 

Many of those cases with the highest uncertainty show severe fat infiltrations of the thigh muscle. \\

\textbf{Alternative reference methods:}

UK Biobank field 23275 contains measurements of the lean mass of the legs by DXA for 5,170 subjects. These values describe more than just muscle volume, but may still be considered as a proxy. These values were first converted from mL to L and then mapped to the target with the following linear transformation parameters:

($0.69x + 0.64 L$).

UK Biobank return 981 by application 23889 also offers thigh muscle volume measurements for 9,441 subjects. These values were first converted from mL to L and then mapped to the target with the following linear transformation parameters:

($1.06x + 0.67 L$).

\begin{table}[H]

\begin{center}	

\renewcommand{\arraystretch}{1.25}	

\caption{Comparison of TTM references}

\label{tab_supp_alt_ttm}

\begin{tabular}{llrrrrr}

\hline	

\hline	

Method & N &  ICC & R2 & MAE & MAPE & r\\

\hline			

Proposed & 4,483 & \textbf{0.996} & \textbf{0.992} & \textbf{0.173} & \textbf{1.7} & \textbf{0.997}  \\

Field 23275 & 4,483 & 0.958 & 0.919 & 0.561 & 5.6 & 0.959 \\

\hline

Proposed & 8,144 & \textbf{0.997} & \textbf{0.993} & \textbf{0.161} & \textbf{1.6} & \textbf{0.997} \\

Return 981 & 8,144 & 0.989 & 0.978 & 0.284 & 2.8 & 0.989 \\			

\hline	

\hline	

\multicolumn{7}{l}{*Comparison to the target values, listing both the proposed predictions}\\  \multicolumn{7}{l}{  and alternative UK Biobank reference values on the same subjects}

\end{tabular}

\end{center}

\end{table}

\newpage

\textbf{Aggregated saliency (TTM):}

\begin{figure}[H]

\centering	

\includegraphics[width=\textwidth]{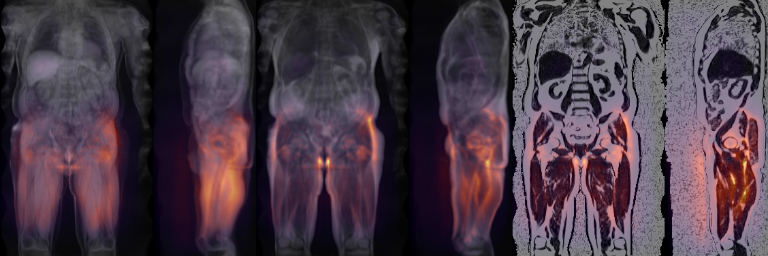}			

\includegraphics[width=\textwidth]{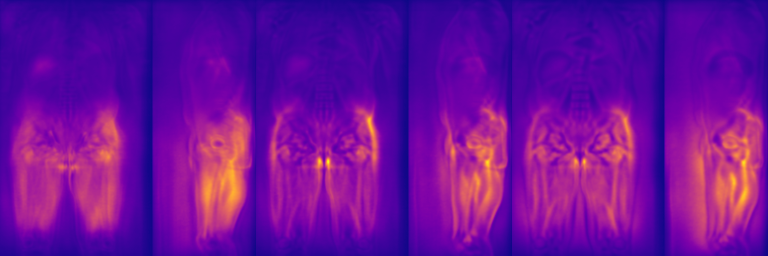}					

\caption{Aggregated saliency \citep{Langner2019} for Total Thigh Muscle (TTM) for 3,091 subjects, generated by a single \textit{mean-variance} network. Each row shows the water, fat, and fat-fraction channels side by side, with the top row showing an overlay on the image data and the bottom row the saliency only.}

\label{supp_fig_sal_ttm}	

\end{figure}

\newpage

%%%%%%%%%%%%%%%%%%%%%%%%%%%%%%%%%%%%%%%%%%%%%%%%%%%%%%%%%%%%%%%%%%%%%%%%%%%%%%%%%%%%%%%%%%%%%%%

% LFF

\newpage

\subsubsection{\Large \textbf{Liver Fat Fraction (LFF)}}

\begin{figure}[H]

\centering	

\includegraphics[width=\textwidth]{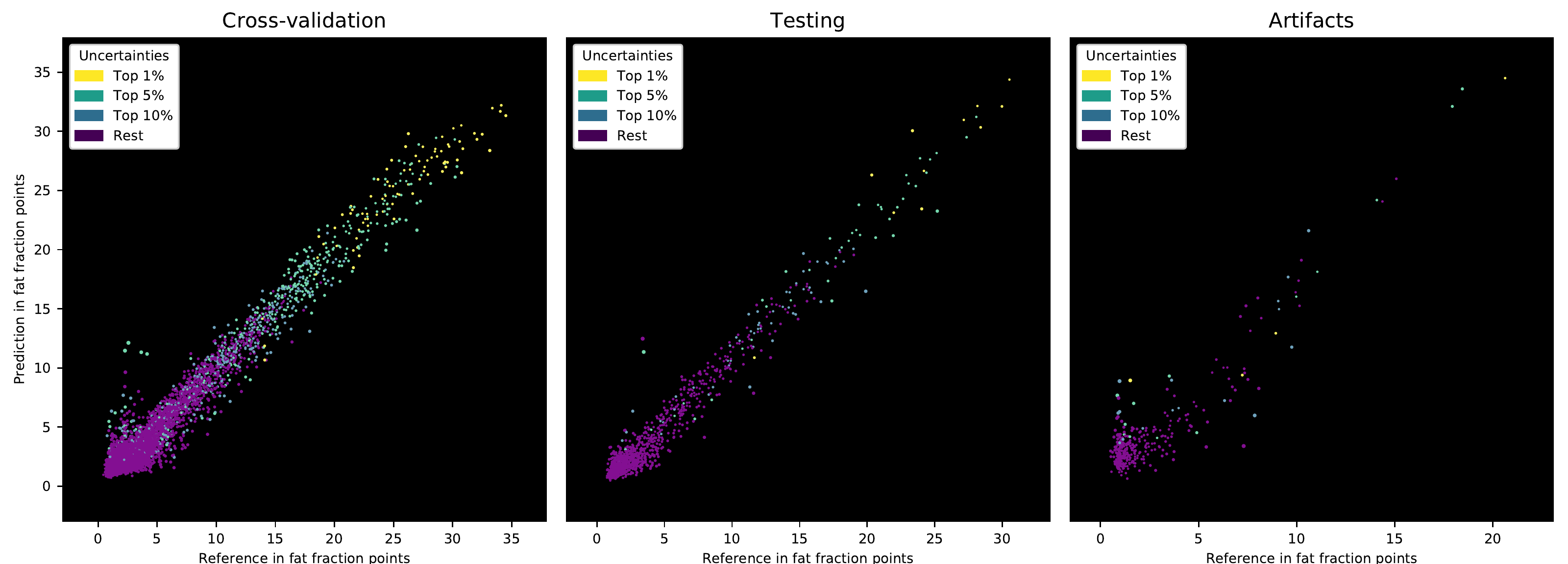}		

\caption{Predictions in cross-validation on \textit{$D_{cv}$}, testing on \textit{$D_{test}$}, and on subjects with artifacts of \textit{$D_{art}$}, with color-coded uncertainty. }

\label{supp_fig_scatter_lff}	

\end{figure}

\begin{figure}[H]

\centering	

\includegraphics[width=\textwidth]{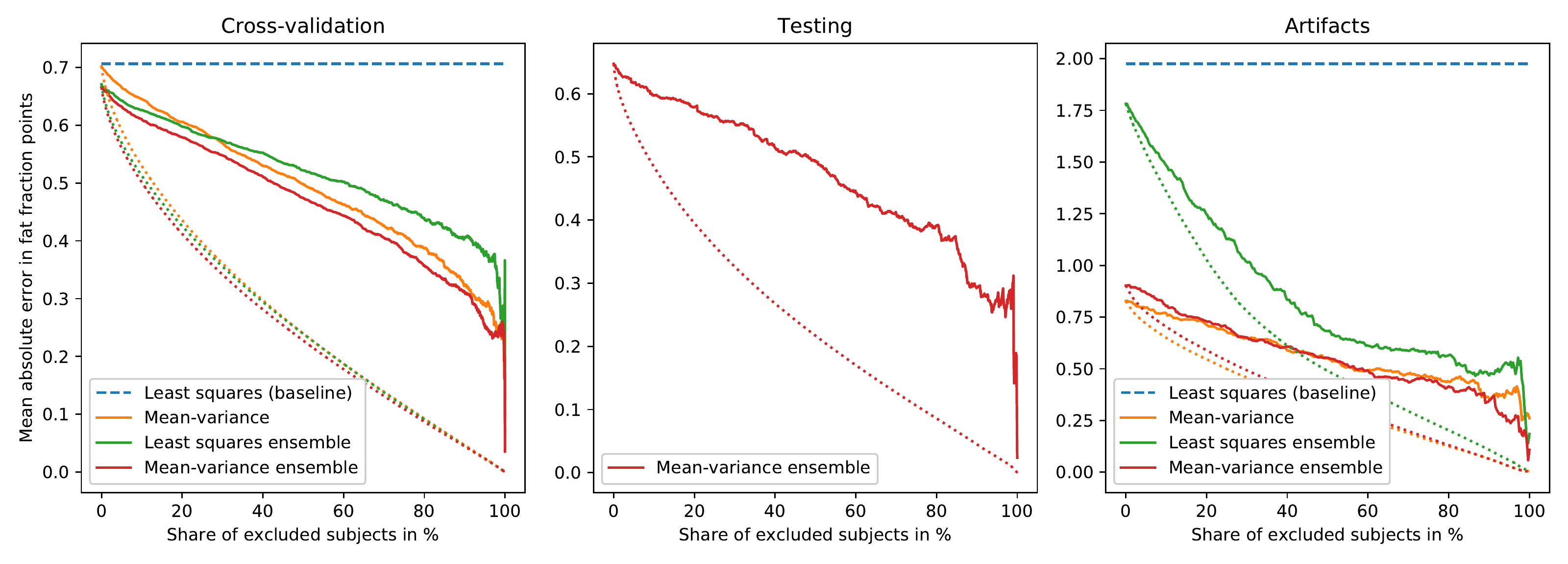}		

\caption{Sparsification in cross-validation on \textit{$D_{cv}$}, testing on \textit{$D_{test}$}, and on subjects with artifacts of \textit{$D_{art}$}, with oracle curves (dotted).}

\label{supp_fig_sparsify_lff}	

\end{figure}

\begin{figure}[H]

\centering	

\includegraphics[width=\textwidth]{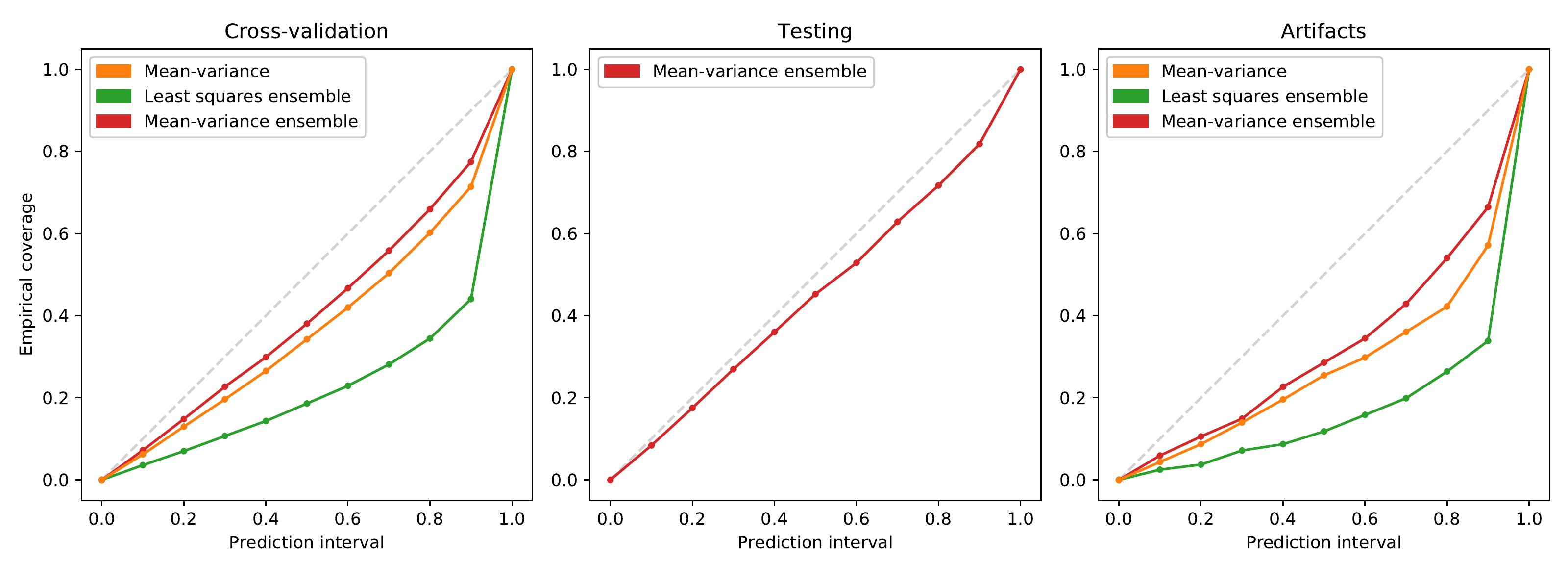}		

\caption{Calibration in cross-validation on \textit{$D_{cv}$}, testing on \textit{$D_{test}$}, and on subjects with artifacts of \textit{$D_{art}$}.}

\label{supp_fig_calibration_lff}	

\end{figure}

\newpage

\textbf{Liver Fat Fraction (LFF)}, extended notes:

The scatter plots of Supplementary Fig. \ref{supp_fig_scatter_lff} show that a small number of samples in the range of zero to five fat fraction points are severely overestimated, both in cross-validation and testing. Not all of these predictions incur high uncertainty.

\hspace{1em}Visual control of the affected subjects showed that the predictions by the proposed method often provided a better match to the neck-to-knee body MRI than achieved by the reference values. No obvious confounding factors such as artifacts or high liver iron content were observed. A similar effect was noted in previous work \citep{langner2020large} where a \textit{least squares} regression technique was trained to emulate an alternative set of UK Biobank liver fat measurements, field 22402. As both of these reference fields are based on the dedicated liver MRI instead of the neck-to-knee body MRI used here, a possible explanation could be an unusually severe mismatch of both protocols for these subjects.

\hspace{1em}On average, LFF incurred by far the highest normalized uncertainties (calculated by dividing the predicted uncertainty by the predicted means) of all targets. Finally, it is worth noting that for this target superior results may be possible when using an input format that only shows a fat fraction slice of the upper body, as previously proposed \citep{langner2020large}, although no rigorous comparison was attempted in the scope of this work. The technique could also be applied directly to the dedicated liver MRI. \\

\textbf{Alternative reference methods:}

UK Biobank field 22402 contains alternative liver fat fraction values for 4,616 subjects, obtained by mostly manual analysis of dedicated liver MRI \citep{wilman2017characterisation}. Relative to the target used in this work, one outlier subject is overestimated by 24 fat fraction points and no linear transformation was applied.

\begin{table}[H]

\begin{center}	

\renewcommand{\arraystretch}{1.25}	

\caption{Comparison of LFF references}

\label{tab_supp_alt_lff}

\begin{tabular}{llrrrrr}

\hline	

\hline	

Method & N &  ICC & R2 & MAE & MAPE & r\\

\hline			

Proposed & 4,401 & 0.978 & 0.956 & 0.669 & 26.3 & 0.978 \\			

Field 22402 & 4,401 & \textbf{0.987} & \textbf{0.972} & \textbf{0.430} & \textbf{14.8} & \textbf{0.989} \\

\hline	

\hline	

\multicolumn{7}{l}{*Comparison to the target values, listing both the proposed predictions}\\  \multicolumn{7}{l}{  and alternative UK Biobank reference values on the same subjects}

\end{tabular}

\end{center}

\end{table}

\newpage

\textbf{Aggregated saliency (LFF):}

\begin{figure}[H]

\centering	

\includegraphics[width=\textwidth]{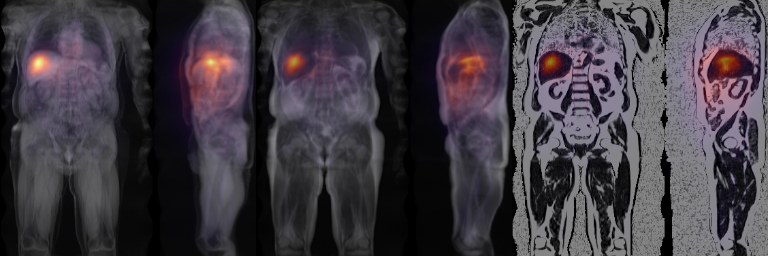}			

\includegraphics[width=\textwidth]{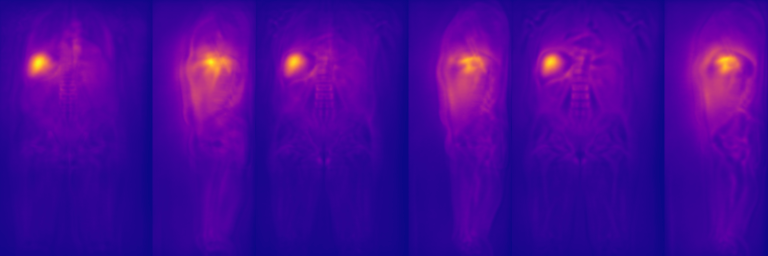}					

\caption{Aggregated saliency \citep{Langner2019} for Liver Fat Fraction (LFF) for 3,091 subjects, generated by a single \textit{mean-variance} network. Each row shows the water, fat, and fat-fraction channels side by side, with the top row showing an overlay on the image data and the bottom row the saliency only.}

\label{supp_fig_sal_lff}	

\end{figure}

\newpage

\subsection{Inference}

\begin{figure}[h]
\centering	
\includegraphics[width=0.45\columnwidth]{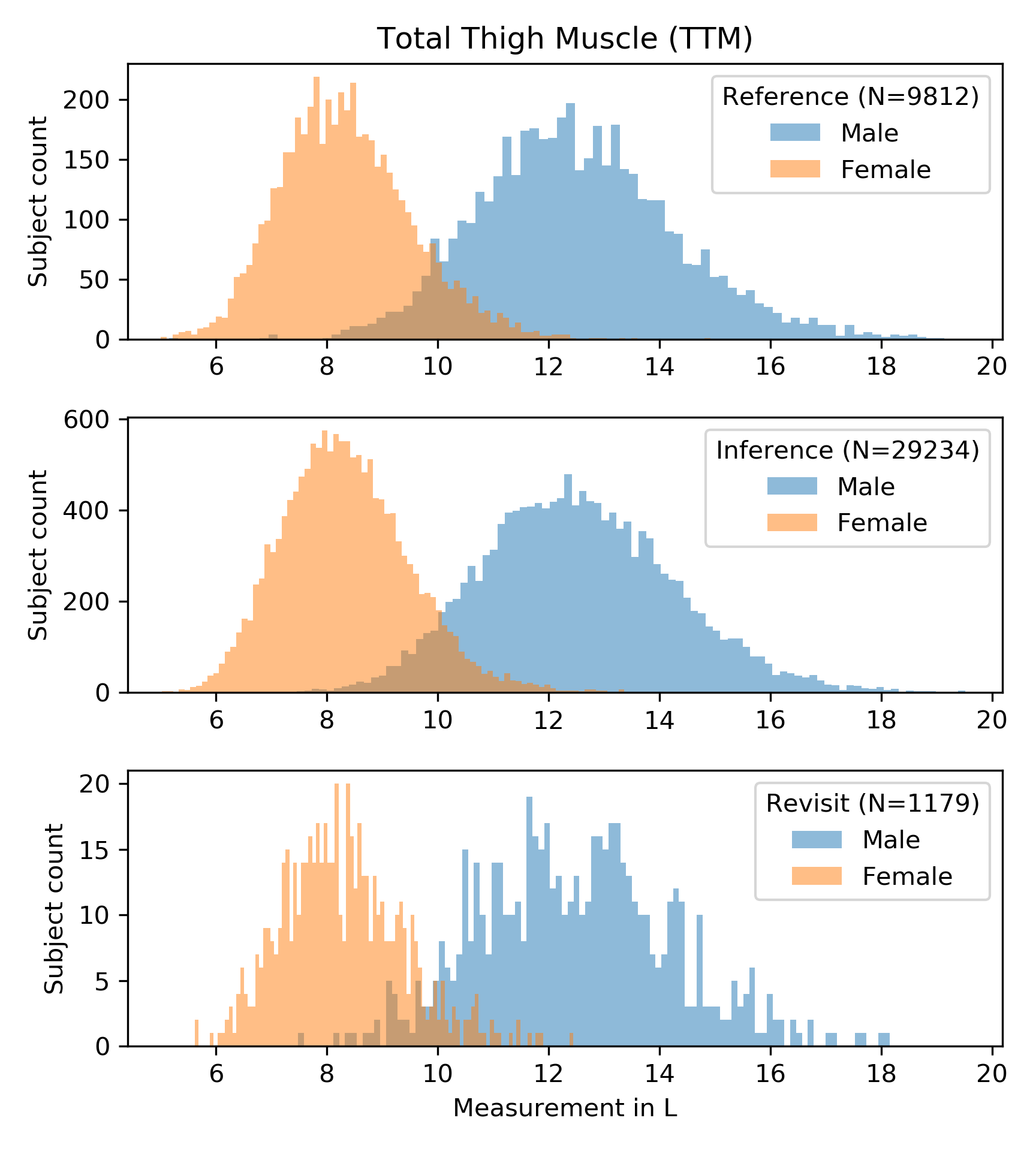}	
\includegraphics[width=0.45\columnwidth]{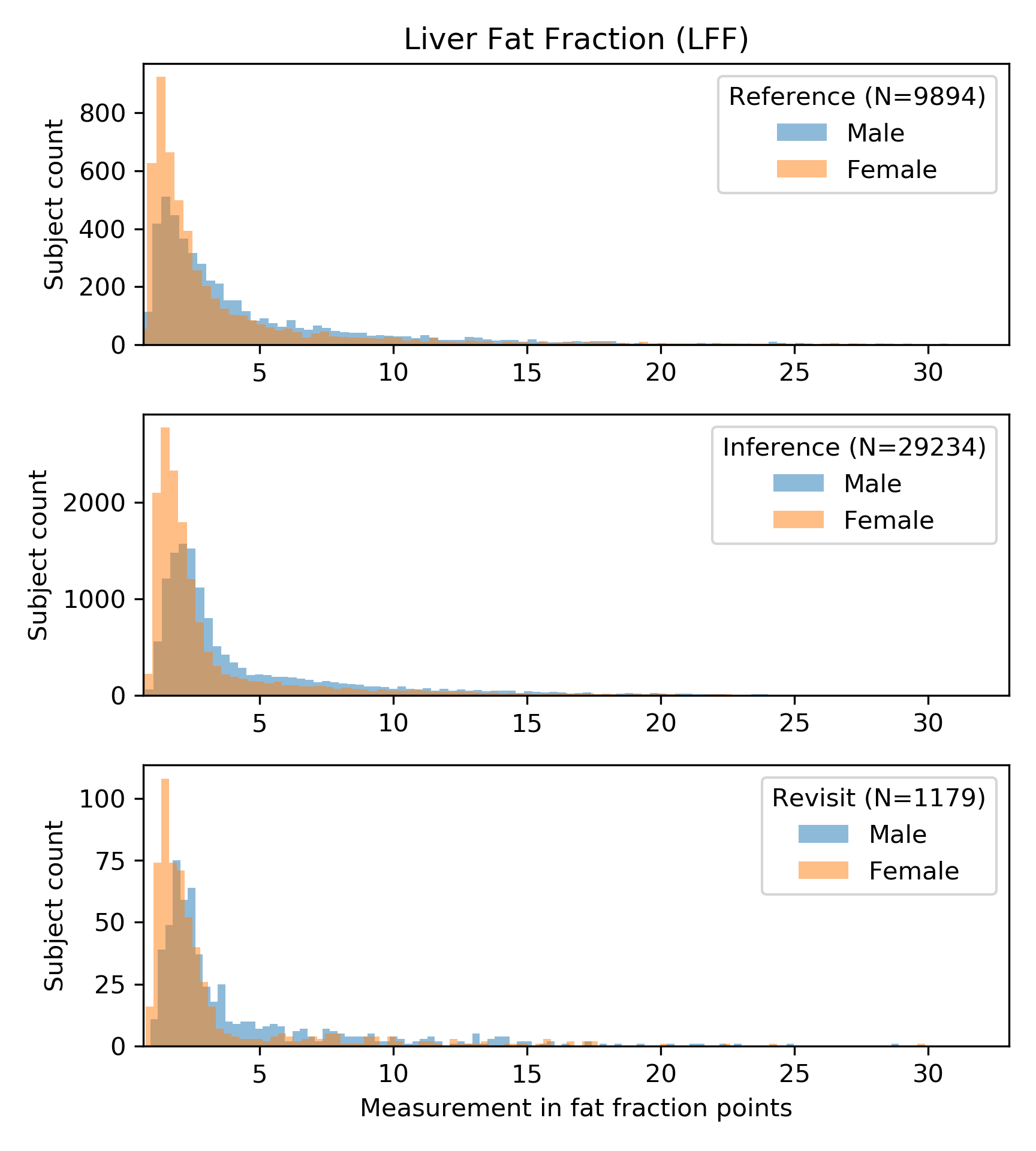}				
\caption{Reference and predicted Total Thigh Muscle (TTM) (left column) and Liver Fat Fraction (LFF)
(right column).}	
\label{supp_fig_hist_ttm_vat}
\end{figure}

The following histograms of Supplementary Fig. \ref{supp_fig_hist_sat_lff}, \ref{supp_fig_hist_tat_tlt}, and \ref{supp_fig_hist_ttm_vat} show the reference values in comparison to those measurements predicted for inference on the original imaging visit on dataset \textit{$D_{infer}$} and the later repeat imaging visit \textit{$D_{revisit}$}. All shown data passed the visual quality controls, but no further attempt was made to exclude outliers based on the predicted uncertainty for these plots.
\vspace{-0.5cm}
\begin{figure}[h]
\centering	
\includegraphics[width=0.45\columnwidth]{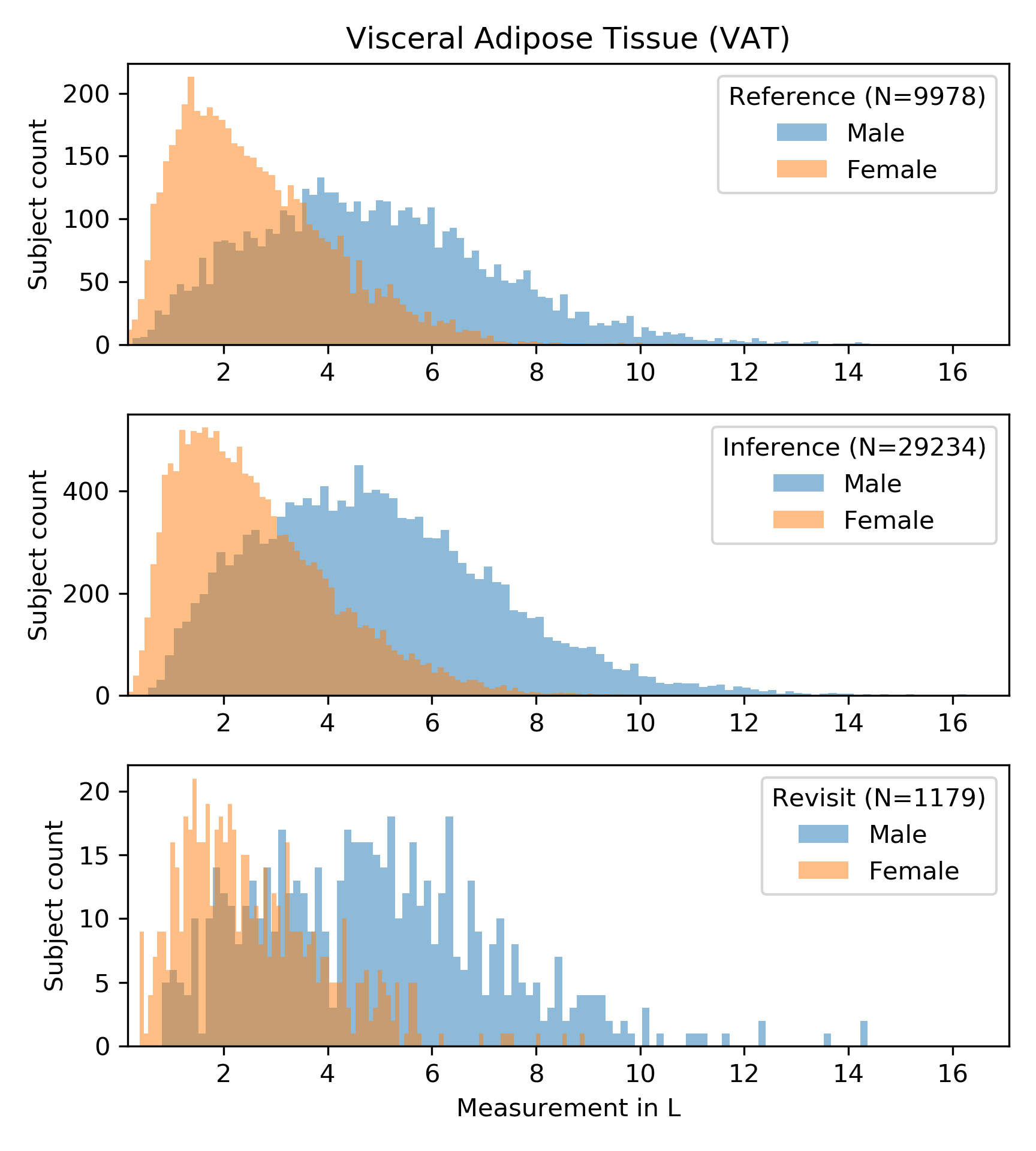}	
\includegraphics[width=0.45\columnwidth]{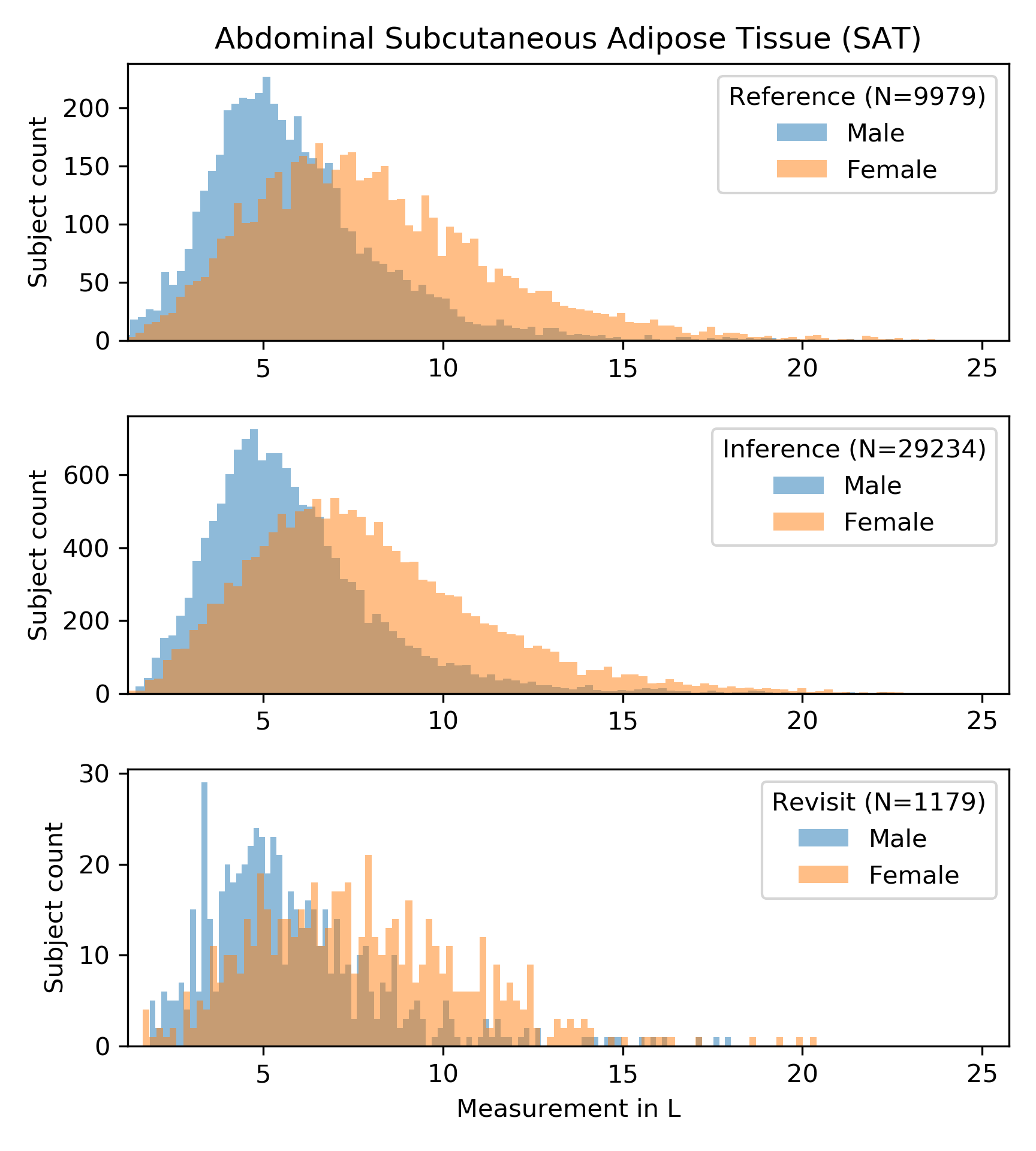}		
\caption{Reference and predicted Visceral Adipose Tissue (VAT) (right column) and Subcutaneous Adipose Tissue (SAT) (right column).}
\label{supp_fig_hist_sat_lff}
\end{figure}
%\vspace{-0.5cm}
\begin{figure}[H]
\centering	
\includegraphics[width=0.45\columnwidth]{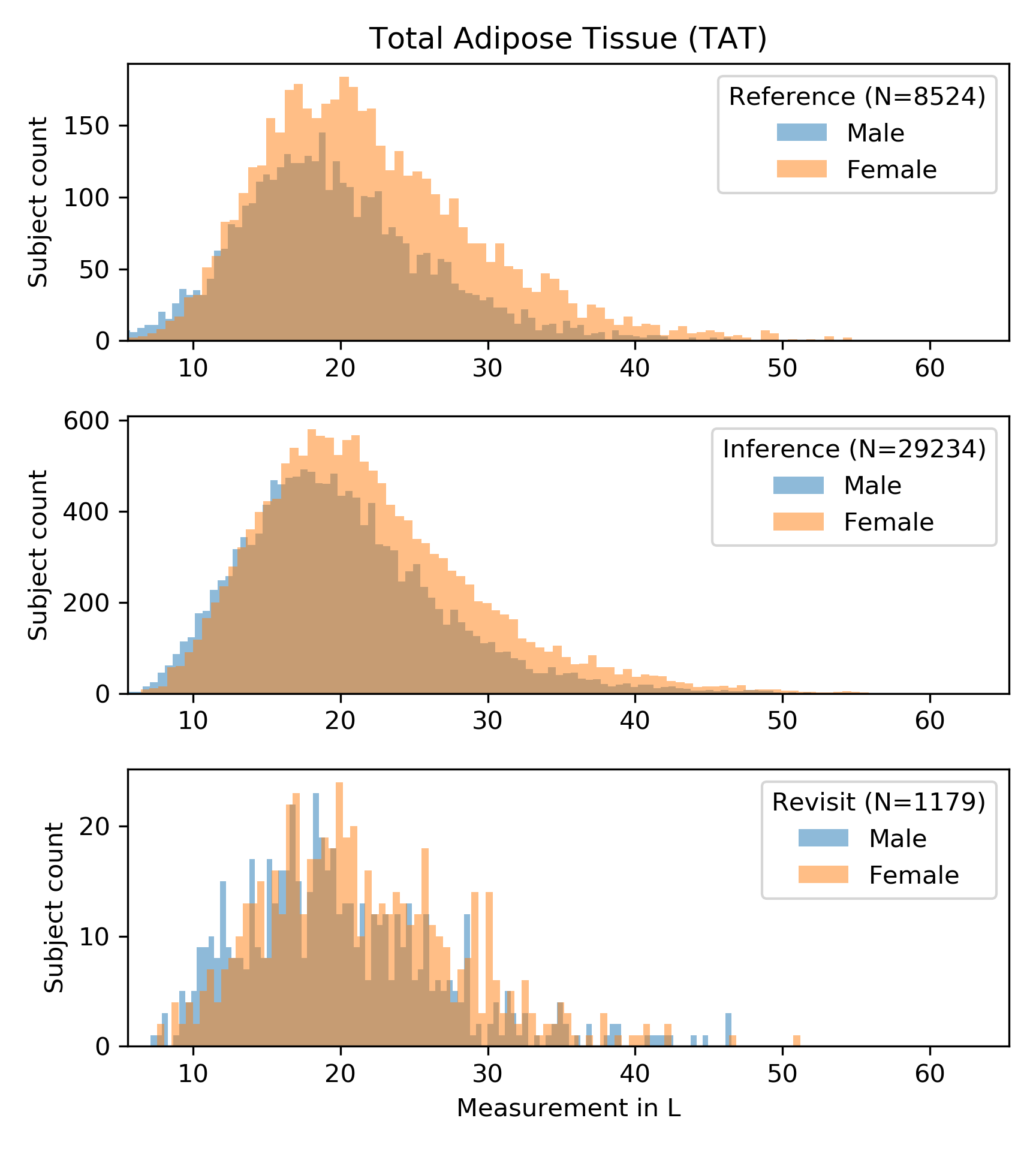}	
\includegraphics[width=0.45\columnwidth]{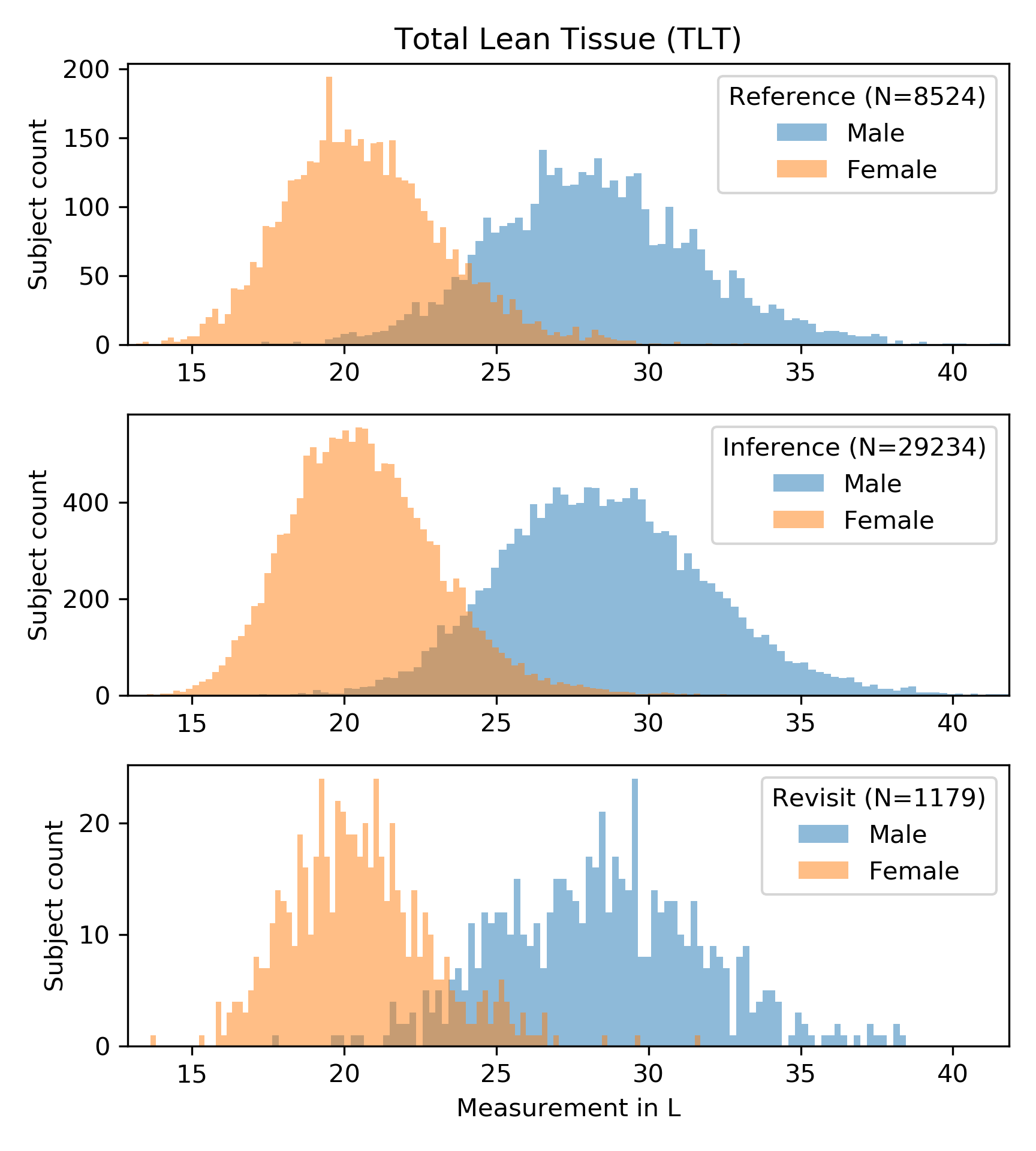}		
\caption{Reference and predicted Total Adipose Tissue (TAT) (left column) and Total Lean Tissue (TLT) (right column).}
\label{supp_fig_hist_tat_tlt}
\end{figure}

\vfill

%\clearpage

\end{document}